\documentclass[showpacs,preprintnumbers,amssymb,twocolum,aps,reprint,superscriptaddress,showkeys,amsmath,floatfix]{revtex4-2}

\usepackage{graphicx}
\usepackage{dcolumn}
\usepackage[dvipsnames]{xcolor}
\usepackage{bm}
\usepackage{amsmath}
\usepackage{amssymb}
\usepackage{epsfig}
\usepackage{amsfonts}
\usepackage{lineno,hyperref}
\usepackage{array}
\usepackage{float}
\usepackage{microtype}
\usepackage{multirow}
\usepackage{adjustbox}
\usepackage[english]{babel}
\usepackage{epstopdf}
\usepackage{blindtext}
\usepackage{booktabs}
\usepackage{subcaption}
\usepackage[a4paper, total={6.5in, 10in}]{geometry}
\usepackage{appendix}

\def \a{\alpha}
\def \b{\beta}

\def \o{\omega}

\def \be{\begin{equation}}
\def \ee{\end{equation}}
\def \ben{\begin{eqnarray}}
\def \een{\end{eqnarray}}

\begin{document}

\title{From Geometry to Observation: Gravitational Waves and the Raychaudhuri Equation}

\author{Sougata Bhunia}
\email{sougatabhunia066@gmail.com}
\affiliation{Department of Physics, Prabhat Kumar College, Contai, Purba Medinipur 721404, India}

\author{Anubhab Dutta}
\email{anubhabduttabnk@gmail.com}
\affiliation{Department of Physics, School of Natural Sciences, Sister Nivedita University, DG 1/2, Action Area 1 Newtown, Kolkata 700156, India.}

\author{Debashis Gangopadhyay}
\email{debashis.g@snuniv.ac.in}
\affiliation{Department of Physics, School of Natural Sciences, Sister Nivedita University, DG 1/2, Action Area 1 Newtown, Kolkata 700156, India.}

\author{Goutam Manna$^a$}
\email{goutammanna.pkc@gmail.com \\$^a$Corresponding author}
\affiliation{Department of Physics, Prabhat Kumar College, Contai, Purba Medinipur 721404, India} 
\affiliation{Institute of Astronomy, Space and Earth Science, Kolkata 700054, India}

\date{\today}

\begin{abstract}
Gravitational waves (GWs) are independent of any particular theory of gravity. The universality of this notion is highlighted by the Raychaudhuri equation (RE), which is independent of any theory of gravity and contains the Ricci tensor $R_{\mu\nu}$ as a key ingredient, thereby connecting spacetime geometry with matter-energy content. Under small metric perturbations, $R_{\mu\nu} \propto \Box h_{\mu\nu}$, where $h_{\mu\nu}$ is the perturbation, indicating that various gravity theories, via their corresponding $R_{\mu\nu}$, produce different gravitational wave equations. In the framework of Einstein's gravity, this leads to the standard wave equation. This study analyzes a modified form, {\it GW-inspired RE}, within the homogeneous and isotropic FLRW background to investigate late-time cosmic acceleration and structure formation. We employ {\it Pantheon+ SNe Ia, Hubble, and BAO} datasets to constrain model parameters through Bayesian inference utilizing NUTS in {\it NumPyro}. A nuisance parameter $\mu_0$ is introduced to address residual systematics. This facilitates a robust estimation of $H_0$, $\Omega_{DE,0}$, and $r_d$, which addresses the resolution of the Hubble tension. We analyze the redshift evolution of the deceleration parameter, $q(z)$, both with and without $\mu_0$, emphasizing its influence on cosmic dynamics. The GW-inspired RE is reformulated as a harmonic oscillator, providing insight into expansion and geodesic focusing. A graphical comparison demonstrates the relationship $d^{GW}_L(z) = d^{EM}_L(z)$ utilizing GWOSC data. Thus, the RE in the context of small perturbation of the metric opens up whole new vistas of {\it observational astronomy.} 
\end{abstract}

%\begin{highlights}
%\item The GW-inspired RE connects metric perturbations to the geometry of spacetime in various gravitational theories.
%\item The modified RE in FLRW background models effectively addresses cosmic acceleration and structure formation.
%\item Bayesian inference utilizing datasets constrains essential cosmological parameters and addresses the Hubble tension.
%\item Reformulating RE as a harmonic oscillator elucidates the dynamics of expansion and the convergence of geodesics.
%\end{highlights}

%%%%%%%%%%%%
%\pacs{04.60.−m, 04.70.Dy, 04.62.+v}
%%%%%%%%%%%%%
\maketitle

\section{Introduction} 
The first theoretical description of gravitational waves was presented in Einstein's landmark 1916 article \cite{Einstein}. Subsequent important contributions came in  \cite{Bondi1, Bondi2, Rickles, Mauro, Schutz, Weinberg}.
The Laser Interferometer Gravitational-Wave Observatory (LIGO) \cite{Abbott} first detected gravitational waves, thereby validating one of the most significant predictions of Einstein's general theory of relativity (GTR). The detection originated from the merging of two black holes, about a $1.3$ billion light-years apart, which generated waves in spacetime as they spiraled towards one another and then merged in a catastrophic event. This result not only confirmed the presence of gravitational waves but also provided empirical evidence for the dynamics of binary black hole systems and their merging processes, which had long been theorized but never directly seen previously.

On the other hand, an equation that provides the development of a congruence of geodesics as seen by other neighboring congruences of geodesics was developed in 1955 by A. K. Raychaudhuri \cite{Ray52, Ray53, Ray55, Ray57, Ray, Kar, Ehlers, Ellis, Dadhich}. The author formulated the equation from a purely geometrical standpoint, which is completely independent of any other gravitational theories. 

The relevance of the Raychaudhuri equation (RE) in modern cosmology is many. The RE is used to show that gravity acts between two material things in an attractive manner. The renowned Hawking-Penrose singularity theorem   \cite{Penrose65,Hawking65,Hawking66,Senovilla,Bhattacharyya} was studied using it as one of the basic lemmas. RE takes into account the universe's non-homogeneity (in terms of shear) and non-isotropy (in terms of rotation) \cite{Kar}. RE shows how fast two null or timelike geodesics can grow. Depending on the Ricci tensor's energy conditions, these geodesics may converge or diverge. The focusing theorem is explained by convergent geodesics \cite{Poisson, Das}, while divergent geodesics explains the non-focusing theorem via the universe's expansion \cite{Das, Choudhury}. By examining the behavior of null or timelike geodesics in relation to the expansion or contraction influenced by the energy conditions associated with the Ricci tensor, one can analyze key quantities in a collapsing star \cite{Choudhury2,Hensh}, such as the negative scalar expansion ($\theta$) and the shear tensor ($\sigma_{ab}$), which influence the rate of change of the scalar expansion. The Ricci curvature tensor, related to matter content and energy density, also plays a role in the evolution of $\theta$. The equation is crucial for understanding the conditions under which singularities can form, as it provides a mathematical framework for understanding the dynamics of expansion and contraction in systems like collapsing stars under gravity \cite{Prisco}. The RE can be modified in the presence of scalar fields \cite{Das} and may elucidate the inhomogeneities of the cosmos, which can be correlated with inflationary-type fields \cite{Panda1}. The modified RE may elucidate the gravito-quantum effects via the phase of the universe characterized by the violation of the null energy condition (NEC) \cite{Panda2}.

In this article, we show that the presence of gravitational waves is already inherent in the RE, which is independent of any theory of gravity. This is because of the presence of the Ricci tensor as a key ingredient of the RE. We work in the context of linearized gravity and formulate a localized energy-momentum connection that allows us to reformulate the RE into a localized version.  We shall show in the context of small metric perturbations that the Ricci tensor is proportional to the metric perturbations. Therefore, as different theories of gravity will give different equations of motion involving the Ricci tensor, the relevant wave equations satisfied by the metric perturbations will also be different. In the context of Einstein's gravity, we will get the usual gravitational wave equations. However, for other different theories of gravity, the field equations involving the Ricci tensor would be different, and hence the relevant wave equations for metric perturbations will also be different. This will give different solutions for gravitational waves. This possibility reveals the fundamental geometric nature of the RE, which enables one to distinguish between different theories of gravity at the level of cosmological observations. {\it Therefore, the RE in the context of small metric perturbation opens up new vistas of observational astronomy.} In this context, our work demonstrates that the GW-inspired RE, when applied to Friedmann–Robertson–Walker (FLRW) spacetime, may provide observationally meaningful insights into gravitational waves.

We organize the paper as follows: Section II provides a brief discussion of the RE. Section III presents the Raychaudhuri equation in the context of small metric perturbations based on the relationship between the Ricci tensor and the perturbative metric tensor. This section also explores the localized form of the Raychaudhuri equation using metric perturbations. Section IV extends that RE in the background of FLRW universe and we have done a data analysis to fit our model against $PANTHEON+SHOES$ data, {\it {the Observational Hubble Dataset (OHD)}} and {\it{BAO}} data.  Section V concludes with a summary of our findings.

\section{The Raychaudhuri Equation } 
Raychaudhuri analyzed trajectories of congruence of geodesics along time-like or null geodesics, which characterize the motion of observers or test particles traversing space-time only under gravitational effect \cite{Ray55, Ray}. Observers traversing time-like geodesics may be conceptualized as stationary in relation to a comoving tetrad, which is a locally specified coordinate system that moves with the observer. These geodesics represent the paths of free-falling motion characterized by a velocity $u^{\alpha}$. A time-like geodesic flow may be like the movement of fluid along streamlines \cite{Landau}. The ideas of divergence, shear, and vorticity are used to describe geodesic congruence. Subsequent authors expanded Raychaudhuri’s approach by including accelerated motion when the velocity field $u^{\alpha}$ exhibits a non-zero acceleration $\Dot{u}^{\alpha}$ where the change is with respect to an affine parameter $\lambda$  parametrizing the geodesics. 

However, it should be noted that the geodesic equation, geodesic deviation equation, and Raychaudhuri equation are not the same when studying cosmic or gravity situations. The difference may be described as follows: The geodesic equation represents the trajectory of an individual particle or light beam in spacetime, adding spacetime curvature using Christoffel symbols. It is used to investigate free particle motion and gravitational lensing. The geodesic deviation equation, which involves the Riemann tensor, investigates the relative motion of neighboring geodesics, taking into account tidal effects, gravity waves, and geodesic stability. The Raychaudhuri equation examines the evolution of a family (congruence) of geodesics, concentrating on expansion, shear, and vorticity, and has applications in gravitational focusing, singularity theorems, and cosmology. Here, we are concerned about the Raychaudhuri equation only for our study not concerned about the geodesic equation and geodesic deviation equation.

Caustics are regions of spacetime where a congruence of geodesics, or a collection of closely aligned geodesics, converges, resulting in their intersection or near intersection. This usually occurs because of variations in spacetime curvature resulting in the convergence of geodesics. The production of caustics fundamentally arises from the geometric behavior of geodesics, which create conjugate points or meet, with these intersections often associated with singularities in the congruence's expansion or focusing characteristics. Physically, this change corresponds to a variation in volume while conserving mass, as dictated by the equation of continuity \cite{Ray55, Ray, Kar, Ehlers, Ellis, Dadhich}.

Taking into account a representative length as a function of coordinates $\ell(x^{\alpha})$, the comoving volume scales as $\ell^{3}$. Thus, the volume's rate of change $\Theta$ (influenced by anisotropy and vorticity) is \cite{Ellis}: 
\ben
\Theta = \nabla_{\alpha}u^{\alpha} = 3 \dfrac{1}{\ell} \dfrac{d\ell}{d\lambda}
\label{1}
\een.

Here anisotropy means directional variations in fluid flow velocity arising from a shearing force applied perpendicular to the flow direction modifying the fluid's cross-sectional area. Mathematically, this is the symmetric component of the velocity gradient tensor \cite{Kar}. Including contributions from acceleration, we have \cite{Dadhich}:
\ben 
\sigma_{\alpha\beta}=\dfrac{1}{2}(\nabla_{\beta}u_{\alpha}+\nabla_{\alpha}u_{\beta})-\dfrac{1}{3}\Delta_{\alpha\beta}\Theta- \dot{u}_{(\alpha}u_{\beta)}
\label{2}
\een
where $\sigma_{\alpha\beta}$ represents the shear tensor;
 $\Delta_{\alpha\beta}= g_{\alpha\beta} + u_{\alpha}u_{\beta}$ ; with $g_{\alpha\beta}$ the metric tensor. $\Delta_{\alpha\beta}$ projects $\Theta$ onto a hypersurface orthogonal to the direction of fluid flow.

A uniform, irrotational fluid flow retains its irrotational nature unless local vorticity is introduced \cite{Landau}. For such flows, considering a closed contour $C$, the velocity circulation around the contour is  $\Gamma=\oint \vec{u}\cdot \vec{dl}=0$ where $\vec{dl}$ represents the infinitesimal line element along the closed contour. However, a local disturbance can lead to the formation of non-zero vorticity, and 
$\Gamma=\oint_{\delta C}\vec{u}\cdot\vec{dl}
=\oint_{\delta f}(\nabla \times \vec{u})\cdot \vec{df}
\neq 0$. $\delta C$ is the infinitesimal closed contour surrounding the disturbed region and $\delta f$ is the surface element enclosed by, $\delta C$ and we have used Stokes theorem. $\nabla \times \vec{u}$ indicates the vorticity at the location of disturbance.

Raychaudhuri defined vorticity as a spin \cite{Ray55, Kar, Dadhich}, represented as a skew-symmetric second-rank tensor, the antisymmetric component of the velocity gradient. This is articulated alongside the acceleration term as follows: \ben 
\omega_{\alpha\beta}=\dfrac{1}{2}(\nabla_{\beta}u_{\alpha}-\nabla_{\alpha}u_{\beta})- \dot{u}_{[\alpha}u_{\beta]}.
\label{3} 
\een

The velocity gradient can be expressed as follows \cite{Kar, Dadhich}: 
\ben 
\nabla_{\beta}u_{\alpha} = \sigma_{\alpha\beta} + \omega_{\alpha\beta} + \dfrac{1}{3}\Delta_{\alpha\beta}\Theta + \dot{u}_{\alpha}u_{\beta}. 
\label{4} 
\een
The process of parallel transporting the velocity vector around a closed loop on a curved manifold results in a deviation from its initial configuration and it follows that \cite{Ray55, Kar, Dadhich}:
\ben [\nabla_{\beta},\nabla_{\gamma}]u^{\alpha} = R^{\alpha}_{~\delta\beta\gamma}u^{\delta}, \label{5} 
\een
where $R^{\alpha}_{~\delta\beta\gamma}$ denotes the Riemann curvature tensor.
Using all of the above one arrives at the RE for time-like geodesics:
\ben 
\dfrac{d\Theta}{d\lambda}+ 2(\sigma^{2}-\omega^{2})+ \dfrac{1}{3}\Theta^{2} -\nabla_{\alpha}\dot{u}^{\alpha}= - R_{\alpha\beta}u^{\alpha}u^{\beta},
\label{6} 
\een
where $2\sigma^{2}=\sigma_{\a\b}\sigma^{\a\b}$, $2\o^{2}=\o_{\a\b}\o^{\a\b}$, and $R_{\a\b}$ is the Ricci tensor. The null geodesic form of the RE can also be written as \cite{Kar} 
\ben 
\dfrac{d\Theta}{d\lambda}+ 2(\sigma^{2}-\omega^{2})+ \dfrac{1}{2}\Theta^{2} -\nabla_{\alpha}\dot{u}^{\alpha}= - R_{\alpha\beta}k^{\alpha}k^{\beta},
\label{7} 
\een
where $k^{\a}$ is the null vector field. It should be noted that in this work, we can not be concerned about the null form of the RE.

\section{Raychaudhuri equation for small metric perturbations}
Weak gravitational fields imply that curvature is minimal. 
This means \cite{Schutz, Weinberg, Maggiore}
\ben 
g_{\a\b}=\eta_{\a\b}+h_{\a\b} 
\label{8}
\een 
where $\eta_{\a\b}$ is the Minkowski metric of flat spacetime and $h_{\a\b}$ is a small perturbation with $||h_{\a\b}||<<1$. 
Here $\eta_{\a\b}$ is used for raising and lowering indices. This framework posits that the observer is situated in the far field or radiation zone where only minor deviations from flat spacetime remain observable. 

The Ricci tensor now becomes \cite{Schutz, Weinberg}:
\ben
R_{\alpha\beta}\approx R_{\a\b}^{(1)}\equiv\dfrac{1}{2}\left(\partial_{\rho}\partial_{\alpha}h^{\rho}_{\beta}+\partial_{\rho}\partial_{\beta}h^{\rho}_{\alpha}-\Box h_{\alpha\beta}-\partial_{\beta}\partial_{\alpha}h \right)\nonumber\\
\label{9}
\een
where $\square=\eta^{\a\b}\partial_{\a}\partial_{\b}$ is the d'Alembert operator, and $h=\eta^{\a\b}h_{\a\b}$ is the trace of the perturbation.
To further simplify the Eq. (\ref{9}), we introduce harmonic coordinates \cite{Weinberg}:
\ben
\overline{h}_{\alpha\beta}= h_{\alpha\beta} - \dfrac{1}{2} \eta_{\alpha\beta}h
\label{10}
\een
Note that $\overline{h}=\eta^{\a\b}\overline{h}_{\a\b}$. Based on the aforementioned definitions, we can write $h=h^{\a}_{\a}=\eta^{\alpha\beta}h_{\alpha\beta}=\eta^{\alpha\beta}\overline{ h}_{\alpha\beta} - \dfrac{1}{2} \eta^{\alpha\beta}\eta_{\alpha\beta}\overline{h}=-\overline{h}$. Similarly, the mixed component is :  $h^{\rho}_{\beta}=\eta^{\rho\alpha}h_{\alpha\beta}=\eta^{\rho\alpha}\overline{h}_{\alpha\beta} - \dfrac{1}{2} \eta^{\rho\alpha}\eta_{\alpha\beta}\overline{h}=\overline{h}^{\rho}_{\beta}-\dfrac{1}{2}\delta^{\rho}_{\beta}\overline{h}$.

It is straightforward to verify that 
\ben
\partial_{\alpha}\overline{h}_{\alpha\beta}=0
\label{11}
\een
i.e. the theory is now in terms of gauge invariant transverse (i.e. physical) variables \cite{Gangopadhyay}.

 Eq. (\ref{9}) thus becomes 	
\ben
 R_{\alpha\beta}= \dfrac{1}{2}\Big(\partial_{\rho}\partial_{\alpha}\overline{h}^{\rho}_{\beta}+\partial_{\rho}\partial_{\beta}\overline{h}^{\rho}_{\alpha}-\Box \overline{h}_{\alpha\beta}+\dfrac{1}{2}\eta_{\alpha\beta}\Box \overline{ h}\Big).\nonumber\\
 \label{12}
\een 

Note that {\it we have not assumed any underlying theory of gravity in arriving at our result}. In Refs. \cite{Schutz, Misner} harmonic coordinates and the relevant gauge transformations are discussed in some detail in the context of Einstein gravity for $g_{\a\b}=\eta_{\a\b}+h_{\a\b}$.  

Implementing Eq. (\ref{10}) in Eq. (\ref{12}) gives the expression for the Ricci tensor as: 
\ben
R_{\alpha\beta}&&= \dfrac{1}{2}\left(-\Box \overline{h}_{\alpha\beta}+\dfrac{1}{2}\eta_{\alpha\beta}\Box \overline{ h}\right)\nonumber\\&&
 = - \dfrac{1}{2}\Box\left(\overline{h}_{\alpha\beta}-\dfrac{1}{2}\eta_{\alpha\beta}\overline{ h}\right)\nonumber\\ &&
= - \dfrac{1}{2}\Box{h_{\alpha\beta}}
\label{13}
\een
The above Eq. (\ref{13}) indicates that within the weak-field approximation and adhering to the Lorenz gauge, {\it the Ricci tensor exhibits a direct proportionality to the wave operator applied to the metric perturbation.} 

Applying this result (\ref{13}) to the Raychaudhuri identity in Eq. (\ref{6}), we get
\ben
\dfrac{d\Theta}{d\lambda}+\dfrac{1}{3}\Theta^{2}+2(\sigma^{2}-\omega^{2})-\nabla_{\alpha}\dot{u}^{\alpha}=\dfrac{1}{2}\Box h_{\alpha\beta}u^{\alpha}u^{\beta}\nonumber\\
\label{14}
\een
This identity (\ref{14}) captures the dynamic nature of spacetime, clearly illustrating {\it the interaction between the kinematical characteristics of geodesic congruences, namely expansion, shear, and vorticity, and the perturbations in the spacetime metric denoted by $h_{\a\b}$.} The above Eq. (\ref{14}), we may also define as a {\it GW-inspired RE.} The inclusion of the term $\Box h_{\alpha\beta}u^{\alpha}u^{\beta}$ on the right-hand side indicates that fluctuations in the spacetime metric tensor play a role in the dynamics of congruences of geodesics, regardless of the particular gravitational framework under investigation, i.e. independent of any particular theory of gravitation. It is important to note the Einstein equation was never employed to get equation (\ref{14}). We derived (\ref{14}) through a purely geometrical process integrated with small metric perturbations via the finding (\ref{13}).

We now briefly illustrate how Eq. (\ref{14}) can be used to study gravitational waves in any theory of gravitation. We first start with Einstein's gravity.

\subsection{Raychaudhuri equation in Einstein's theory of gravitation}

This part highlights the RE (\ref{14}) within the framework of weak fields and its relation to Einstein's theory of gravity. We have Einstein's field equations \cite{Misner, Schutz, Weinberg}
\ben
R_{\alpha\beta}-\dfrac{1}{2}R g_{\alpha\beta}=8\pi G T_{\alpha\beta}
\label{15}
\een
where $G$ is the gravitational constant and $T_{\alpha\beta}$ is the energy-momentum tensor. The energy-momentum tensor 
$T_{\alpha\beta}$ encodes information about the flux of the 
$\a-$th component of the four-momentum across a surface of constant $x^{\beta}$. The Ricci tensor can be expressed in terms of the energy-momentum tensor using Eq. (\ref{15}), as follows:
\ben
 R_{\alpha\beta}= 8\pi G \Big(T_{\alpha\beta}-\dfrac{1}{2}g_{\alpha\beta}T\Big)\label{16}
\een
where $T$ is the trace of the energy-momentum tensor.\\

By applying Eq. (\ref{13}), one can derive the subsequent relationship:
\ben
\Box h_{\alpha\beta}=- 16\pi G\Big(T_{\alpha\beta}-\dfrac{1}{2}g_{\alpha\beta}T\Big)\label{17}
\een
This equation characterizes a wave equation for the metric perturbation $h_{\a\b}$, considering the localized energy delineated by the energy-momentum tensor $T_{\a\b}$. The right-hand side represents the source term produced by the matter and energy distribution within spacetime. 

Note that Eq. (\ref{17}) has some resemblance to the basic wave equation (1.24) in Ref. \cite{Maggiore}, which may be stated as follows:
\ben 
\Box h_{\alpha\beta}=-16\pi G T_{\alpha\beta}. \label{18} 
\een
The equations (\ref{17}) and (\ref{18}) differ intrinsically:

The first equation (\ref{17}) emphasizes energy-momentum conservation, with the source term including the trace-adjusted tensor $\Big(T_{\alpha\beta}-\dfrac{1}{2}g_{\alpha\beta}T\Big)$. This is necessary for general relativity consistency. Whereas the second equation (\ref{18}) simplifies the propagation of $h_{\alpha\beta}$ from $T_{\alpha\beta}$, removing the trace correction. Basically, Eq. (\ref{17}) shows the whole structure of small perturbative general relativity using trace-reversed perturbation, assuring higher generality and alignment with Einstein field equations. In contrast, Eq. (\ref{18}) simplifies the formulation for certain cases or approximations, ignoring the trace term.

In the absence of localized energy, where $T_{\a\b}=0$, the Eq. (\ref{17}) reduces to:
\ben
\Box h_{\alpha\beta}=0, 
\label{19}
\een
which corresponds to a homogeneous wave equation. This finding indicates the propagation of free gravitational waves through spacetime, irrespective of any matter sources.

Equation (\ref{17}) indicates the generation of gravitational waves resulting from fluctuations in spacetime induced by energy and matter distributions. These waves transmit energy away from their sources, such as merging binary stars or black holes. This corresponds with the predictions of Einstein's GTR and has been empirically validated by the observation of gravitational waves by facilities such as LIGO and Virgo \cite{Abbott}. 

On the other hand, when matter is absent ($T_{\a\b}=0$), Eq. (\ref{19}) indicates that gravitational waves can propagate through a vacuum of spacetime. These waves are ripples in the curvature of spacetime, propagating at the invariant speed of light, and are derived from the Einstein field equations under vacuum conditions.

This formulation establishes a theoretical foundation for comprehending gravitational waves as solutions to Einstein's equations, connecting the curvature of spacetime with its energy-momentum content \cite{Misner, Schutz}.\\

Therefore, in the context of the above discussions, we can say that Eq. (\ref{13}) is a fundamental equation valid in any theory of gravity in the presence of small perturbations of the metric.

Now combining Eqs. (\ref{14}) and (\ref{17}), we get a version of GW-inspired RE as
\ben
&\dfrac{d\Theta}{d\lambda}+\dfrac{1}{3}\Theta^{2}+2(\sigma^{2}-\omega^{2})-\nabla_{\alpha}\dot{u}^{\alpha}\nonumber\\&=- 8\pi G\Big(T_{\alpha\beta}-\dfrac{1}{2}g_{\alpha\beta}T\Big)u^{\alpha}u^{\beta}
\label{20}
\een
The equation (\ref{20}) illustrates the connection among various physical factors that influence the behavior of geodesic congruences:

(a) The expression $- 8\pi G\Big(T_{\alpha\beta}-\dfrac{1}{2}g_{\alpha\beta}T\Big)u^{\alpha}u^{\beta}$ on the right-hand side denotes the effect of localized energy and matter, as described by the energy-momentum tensor, on the dynamics of geodesic expansion or contraction. A positive value signifies that energy and local anisotropy play a role in contraction.

(b) The $\sigma^2$ term facilitates local contraction via anisotropic deformation, whereas $\omega^2$ functions as a centrifugal force opposing contraction. This dynamic illustrates the interplay between gravitational collapse and rotational effects in spacetime.

(c) During the expansion of the geodesic congruence ($\Theta>0$), the positive spin or vorticity ($\omega^2$) acts as a repulsive force, reducing the impact of anisotropy ($\sigma^2$). In contrast, contraction ($\Theta<0$) is enhanced by anisotropy but mitigated by vorticity, indicating an inherent balancing mechanism in spacetime dynamics \cite{Ray55}.

Thus, Eq. (\ref{20}) through the relation (\ref{17}) may be elucidated as follows: A spacetime background may locally expand due to an oscillating metric tensor and the existence of positive spin or vorticity. Local anisotropy opposes this expansion, resisting the deformation and influencing the dynamics of the geodesic congruence.

\section{GW inspired RE in the background of FLRW Universe and Observational data analysis}
This section examines the analysis of observational data with the alternate GW-inspired RE presented in Eq.~(\ref{20}). This variant of the RE is especially pertinent for examining observational fingerprints in a cosmological framework, particularly inside the Friedmann–Lemaître–Robertson–Walker (FLRW) backgrounds. Consequently, for simplicity, we examine affinely parameterized ($\dot{u}^{\a}=0$) geodesic congruences. Assuming the universe is homogeneous and isotropic, we use Frobenius' theorem~\cite{Poisson, Das} to establish that the shear tensor is zero, namely, $[ 2\sigma^2 = \sigma^{\alpha\beta} \sigma_{\alpha\beta} = 0. ]$
Correspondingly, the rotation tensor becomes zero given these assumptions, resulting in $[ 2\omega^2 = \omega^{\alpha\beta} \omega_{\alpha\beta} = 0. ]$ 

We consider the background flat FLRW metric as
\ben
ds^2=-dt^2+a^{2}(t)\sum_{i=1}^{3} (dx^{i})^{2}
\label{21}
\een
where $a(t)$ is the scale factor that characterizes the expansion of the universe.

Now, for the consideration of a homogeneous and isotropic universe, Eq. (\ref{20}) can be expressed as 
\ben
\dfrac{d\Theta}{d\lambda}+\dfrac{1}{3}\Theta^{2}=- 8\pi G\Big(T_{\alpha\beta}-\dfrac{1}{2}g_{\alpha\beta}T\Big)u^{\alpha}u^{\beta}
\label{22}
\een
Correspondingly, we consider the energy-momentum tensor as a perfect fluid:
\ben
T_{\alpha \beta}=(\rho+P)u_\alpha u_\beta+ Pg_{\alpha \beta} 
\label{23}
\een
with $(-,+,+,+)$ convention, where $u_\alpha=(1,0,0,0)$ is the fluid four-velocity, $\rho$ is the energy density, and $P$ is the pressure. So that we can directly express the RHS of Eq. (\ref{22}) as
\ben
\Big(T_{\alpha \beta}- \frac{1}{2}g_{\alpha \beta}T\Big)u^\alpha u^\beta= \frac{1}{2}(\rho+3P) 
\label{24}
\een
Therefore, the GW-inspired RE (\ref{22}) in the context of the FLRW background becomes
\ben
3\Big(\dfrac{dH}{dt}\Big)+3H^2=- 4\pi G \sum_{i}(\rho_i+3P_i) 
\label{25}
\een
where the affine parameter $\lambda$ is replaced by cosmic time $t$, the Hubble parameter $H=(\dot{a}/a)$, $\Theta=3H$ and $\rho_i$, $P_i$ are the energy density and pressure of the $i-$th component (i.e. matter, radiation, dark energy). 

The covariant conservation of energy-momentum tensor ($\nabla_\mu T^{\mu \nu}=0$) gives us
\ben
\dot{\rho}_i+3H(\rho_i+P_i)=0.
\label{26}
\een
After solving this Eq. (\ref{26}), we get the following solution:
\ben
\rho_i=\rho_{i,0} a^{-3(1+\o_i)} 
\label{27}
\een
where $\rho_{i,0}$ is the present energy density, and $\o_i$ is the equation of state (EoS) parameter of $i-$th component, with the EoS $P_i=\o_i \rho_i$.

For investigations of observational data analysis, we can convert the above alternate RE (\ref{25}) in terms of redshift ($z$) as follows \cite{Weinberg1, Dodelson}:
\ben
\frac{a(t)}{a_0}= \frac{1}{(1+z)}\nonumber\\
H(z)=-\frac{1}{(1+z)}\dfrac{dz}{dt}
\label{28}
\een
where $a_0$ is the present value of the scale factor.

The standard cosmological model posits that the universe's total energy density consists mostly of three components: dark matter, dark energy, and radiation (including photons and relativistic neutrinos). However, in the present epoch, the contribution from radiation is negligible compared to that of dark matter and dark energy \cite{Planck1}. So we can write 
\ben
\sum_{i=1}^{2} \rho_i=\Big(\rho_{DM}+\rho_{DE}\Big) \label{29}
\een
where $\rho_{DM}$ and $\rho_{DE}$ are the energy density of dark matter and dark energy, respectively.

In the conventional $\Lambda$CDM cosmological model, cold dark matter (CDM) is considered as a pressureless fluid, meaning its pressure $P_{\text{DM}} = 0$. According to the EoS, it follows that $\o_{\text{DM}} = 0$. Conversely, dark energy, represented as a cosmological constant $\Lambda$, has an equation of state parameter $\o_{\text{DE}} = -1$, indicating a constant energy density accompanied by negative pressure \cite{Planck1, Weinberg1}. So using Eqs. (\ref{27}), (\ref{28}) and (\ref{29}), we have
\ben
\rho_{DM}= \rho_{DM,0}(1+z)^3\nonumber\\
\rho_{DE}=\rho_{DE,0}=\text{constant}
\label{30}
\een
where $\rho_{\text{DM},0}$ and $\rho_{\text{DE},0}$ denote the present-day energy densities of dark matter and dark energy, respectively.

Now applying the Eqs. (\ref{28}) and  (\ref{30}) in Eq. (\ref{25}), we get the GW-inspired RE as
\ben
&&\Big(\dfrac{dH}{dz}\Big)=\nonumber\\&& \frac{H}{(1+z)}+\frac{H^2_0}{2H(1+z)} \Big[\Omega_{DM,0}(1+z)^3-2 \Omega_{DE,0}\Big] 
\label{31}
\een
where $H_0 \equiv H~(z=0)$ is the Hubble constant, and $\Omega_{\text{DM},0} = \left( \frac{8\pi G \rho_{\text{DM},0}}{3H_0^2} \right)$ and $\Omega_{\text{DE},0} = \left( \frac{8\pi G \rho_{\text{DE},0}}{3H_0^2} \right)$ represent the present-day dimensionless density parameters for dark matter and dark energy, respectively.

On the other hand, using Eq. (\ref{28}) directly in Eq. (\ref{25}), we can get the following relation:
\ben
\Big(\dfrac{dH}{dz}\Big)= \frac{H}{(1+z)}+\frac{H}{2(1+z)} \Big[\Omega_{DM}-2 \Omega_{DE}\Big] ~~~
\label{32}
\een
By neglecting the radiation-dominated phase of the universe, owing to its minimal impact on observational data analysis, and applying the condition $(\Omega_{DM}(z) + \Omega_{DE}(z) = 1)$, we can completely determine the evolution of the Hubble parameter and the associated energy content of the universe. Therefore, the Eq. (\ref{32}) becomes,
\ben
\Big(\dfrac{dH}{dz}\Big)= \frac{H}{(1+z)}+\frac{H}{2(1+z)} \Big[1-3 \Omega_{DE}\Big] \label{33}
\een
Using the relations $\dot{z} = -H(1+z)$ and $\dot{H} = -H(1+z)\frac{dH}{dz}$, we derive a differential equation for the rate of change of the dark energy density parameter $(\Omega_{DE})$ concerning redshift $z$ as
\ben
\Big(\frac{d\Omega_{DE}}{dz}\Big)=-\frac{2\Omega_{DE}}{H}\Big(\frac{dH}{dz}\Big).
\label{34}
\een

The GW-inspired REs in Eqs.~(\ref{31}), (\ref{32}) and (\ref{33}), constructed inside the FLRW framework, govern cosmological evolution and include essential parameters such as $\Omega_{\text{DE},0}$ and $H_0$, facilitating the analysis of the universe's dynamics influenced by dark matter and dark energy.

\subsection{Model fitting with Observational Data Analysis}

In this subsection, we determine two key cosmological model parameters: the current Hubble constant ($H_0$) and the current dark energy density parameter ($\Omega_{DE,0}$). These values are derived by solving the differential equations (\ref{33}) and (\ref{34}) with suitable initial conditions. The solutions are then fitted to several observational datasets, including Type Ia supernovae from the \textit{Pantheon+SHOES} compilation \cite{Brout, Scolnic}, observational Hubble data (OHD, listed in Table \ref{tab:I}) \cite{zhang, Jimenez, Simon, Stern, Moresco, Hui, Abbott1, Xu, Metin, Eric, Julian, Blake, Betoule, Beutler1, Beutler2, Samushia}, and baryon acoustic oscillation (BAO) measurements \cite{Alam, Eisenstein1, Percival, Glazebrook, Hu, Eisenstein2, Dawson, DESI}. We use the chi-squared ($\chi^2$) statistic for controlling the model parameters, which quantifies the discrepancy between theoretical predictions and observational data, weighted by the corresponding measurement uncertainties. The optimal parameter values are determined by minimizing the $(\chi^2)$ function, hence determining the best-fit scenario that most closely corresponds with the data. Within the context of Bayesian data analysis, the $(\chi^2)$ statistic is intrinsically linked to the likelihood function via the conventional relationship for Gaussian-distributed errors:
$\mathcal{L}(p) \propto \exp\left(-\frac{1}{2}\chi^2(p)\right)$, where $p$ is the set of model parameters. This relationship supports the use of $(\chi^2)$ minimization as an alternative method to Maximum Likelihood Estimation (MLE). Moreover, it is the basis for developing posterior probability distributions in Bayesian parameter inference. Prominent Bayesian inference tools like \textit{emcee}, \textit{PyMC}, and \textit{Stan} use this log-likelihood formulation for models that assume Gaussian noise.

The \textit{Pantheon+SHOES} dataset consists of 1701 high-quality light curves associated with 1550 distinct Type Ia supernovae (SNe Ia), covering a redshift range from $z=0.00122$ to $z=2.2613$ \cite{Brout, Scolnic}. Cosmological model parameters are restricted by conducting a statistical fit between the observable distance moduli, obtained from supernova photometry, and the theoretically anticipated distance moduli calculated from cosmological models. This comparison underpins parameter inference using likelihood maximization or Bayesian techniques. The distance moduli can be defined as,
\ben
\mu(z,\theta)= (m-M) = 5\, log_{10}(d_L(z))+25 \label{35}
\een
where `$m$' is apparent magnitude, which is how bright the object appears from Earth, and `$M$' is absolute magnitude, which is how bright the object would appear if it were at a standard distance of $10$ parsecs and $d_L(z)$ is dimensionless luminosity distance, defined as \cite{Weinberg1}
\ben
d_L(z)=c(1+z) \int_{0}^{Z} \frac{dz^\prime}{H(z^\prime)} \label{36}
\een
where $c$ is the speed of light in unit of $km/s$. Now taking the derivative concerning $z$ in Eq. (\ref{36}), we can get the following differential equation:
\ben
\frac{dd_L(z)}{dz}= \frac{d_L(z)}{1+z}+\frac{c(1+z)}{H(z)}.
\label{37}
\een

The BAO serves as a reliable standard ruler in contemporary observational cosmology, providing precise limitations on the universe's expansion history. The analysis of BAO characteristics in large-scale structure studies, including galaxy clustering mappings, requires the use of numerous essential cosmological distance metrics. The transverse comoving distance $D_M(z)$, the Hubble distance $D_H(z)$, and the volume-averaged distance $D_V(z)$ are crucial for transforming angular and redshift separations into comoving physical scales. These distance metrics allow the determination of essential cosmological parameters, such as the current Hubble parameter $H_0$ and the redshift-dependent dark energy EoS parameter $\omega(z)$, hence offering a crucial evaluation of $\Lambda$CDM and other cosmological models.

In spectroscopic galaxy redshift surveys, the BAO feature appears in both the radial (line-of-sight) and transverse (angular) directions, encapsulating information on cosmic distances and expansion rates. The radial extent of the BAO peak in redshift space, measured by the redshift separation $\Delta z$ related to the sound horizon scale, facilitates a direct calculation of the Hubble parameter $H(z)=\Big( \frac{c \Delta z}{r_d}\Big)$ \cite{Alam, Eisenstein1, Percival, Glazebrook, Ganguly, Ganguly1}. In this context, $r_d$ denotes the comoving sound horizon at the baryon drag epoch. It denotes the greatest distance that sound waves or acoustic waves in photon-baryon plasma might have traversed before the decoupling of photons from baryons in the early cosmos. It can be expressed as \cite{Hu,Eisenstein2},
\ben
r_d= \int^{\infty}_{z_{drag}} \frac{c_s}{H(z)} dz \label{38}
\een
where $z_{drag} (\approx 1020)$ denotes the baryon drag epoch, the redshift at which baryons decouple from the photon-baryon plasma, and $c_s$ represents the sound speed of the photon-baryon fluid. In our study, we consider $r_d$ as a free parameter, enabling it to be directly constrained by observational data. The BAO feature measures the Hubble distance at redshift $z$:
\ben
D_H(z)=\frac{c}{H(z)} 
\label{39}
\een

In the transverse (angular) direction, the BAO scale corresponds to the characteristic angular separation $\Delta\theta=\frac{r_d}{D_M(z)}$, subtended by the comoving sound horizon $r_d$ at given $z$. The estimated angular diameter distance, $D_M(z)$, depends on the expansion history, can be expressed as \cite{Alam, Eisenstein1, Percival, Glazebrook, Ganguly, Ganguly1}:

\ben
D_M(z)= c(1+z) \int^{z}_{0} \frac{dz^\prime}{H(z^\prime)}\label{40}
\een

When considering the cosmological dependence of $r_d$, the principal observables, restricted by BAO observations, are the dimensionless ratios $D_M(z)/r_d$ and $D_H(z)/r_d$. Historically, BAO constraints were represented by a spherically-averaged or isotropically minimized quantity termed $D_V(z)$, defined as \cite{Alam, Eisenstein1, Percival, Glazebrook, Ganguly, Ganguly1}:
\ben
D_V(z)=[zD_M(z)^2 D_H(z)]^\frac{1}{3} \label{41}
\een
The quantity $D_V(z)$ combines the line-of-sight and transverse BAO information into a single scalar observable, making it particularly valuable at low redshifts, where the anisotropy between radial and angular clustering modes is minimal. In such regimes, $D_V(z)$ serves as an efficient proxy for the geometric constraints provided by BAO, facilitating model comparisons and parameter inference in cosmological analyses.

In this analysis, we consider two independent BAO datasets. The first dataset, designated SDSSBAO, consists of $8$ observations of BAO observables obtained from the {\it Sloan Digital Sky Survey (SDSS)}, as documented in Refs. \cite{Alam, Dawson}. The second dataset, referred to as DESBAO, comprises $9$ data points of BAO measurements derived from the {\it Dark Energy Spectroscopic Instrument Data Release 2 (DESI DR2)}, as detailed in Ref. \cite{DESI}. For the purposes of our analysis, we define the combined BAO dataset as the amalgamation of SDSSBAO and DESBAO, jointly referred to as the BAO data. The data sets are shown in Tables \ref{tab:II} and \ref{tab:III}.\\

Now, for the Hubble dataset given in the Table \ref{tab:I} \cite{zhang, Jimenez, Simon, Stern, Moresco, Hui, Abbott1, Xu, Metin, Eric, Julian, Blake, Betoule, Beutler1, Beutler2, Samushia} and for model fitting with model parameters, we compute the $\chi^2$ as \cite{Ganguly, Ganguly1, Hu, Eisenstein1}:
\ben
\chi_H^2= \sum_{i=1}^{44} \Big(\frac{H^{th}(z_i, p)- H^{obs}(z_i)}{\sigma_{H,i}}\Big)^2 \label{42}
\een
where $H^{th}(z_i,p)$ corresponds to the theoretical value that we will get by solving the differential Eq. (\ref{33}), $H^{obs}(z_i)$ stands for the value given in Hubble dataset in Table \ref{tab:I}, and $\sigma_{H,i}$ is the uncertainties in the measurement of $H(z)$ mentioned in that table.\\

The estimation of the $\chi^2$ statistics for the $PANTHEON$ Type Ia Supernova dataset \cite{Brout, Scolnic} necessitates a modified technique owing to the presence of a complete covariance matrix ($C$) of dimensions $1701\times 1701$. This matrix covers the statistical and systematic uncertainty related to the observed distance moduli $\mu(z)$ for all $1701$ light curves. Consequently, the $\chi^2$ is articulated as \cite{Ganguly, Ganguly1},
\ben 
&&\chi^2_{SN}=\nonumber\\&& \Big(\mu_{th}(z_i,p)-\mu_{obs}(z_i)\Big)^T C^{-1} \Big(\mu_{th}(z_i,p)-\mu_{obs}(z_i)\Big)\nonumber\\
\label{43} 
\een
where $\mu_{th}(z_i,p)$ represents the theoretical value derived from Eq. (\ref{35}) by resolving the differential Eqs. (\ref{33}), (\ref{34}), and (\ref{37}) with initial conditions and model parameters as constraints. $\mu_{obs}(z_i)$ represents the observed value found in the $PANTHEON$ dataset \cite{Brout, Scolnic}, while $C^{-1}$ denotes the inverse of the covariance matrix.\\

For the BAO dataset \cite{Hogg, Hogg1}, the $\chi^2$ is defined as \cite{Ganguly, Ganguly1}:
\ben \chi_i^2= \sum_{i=1}^N\Big(\frac{X_i^{th}(z,p)-X_I^{obs}(z_i)}{\sigma_i}\Big)
\label{44}
\een
where $X_i^{th}(z_i, p)$ represents either $\frac{D_M}{r_d}$, $\frac{D_H}{r_d}$, or $\frac{D_V}{r_d}$, computed theoretically through model fitting. Consequently, the aggregate $\chi_t^2$ associated with the BAO dataset is articulated as: $\chi_t^2=[\chi^2_{(\frac{D_M}{r_d})}+\chi^2_{(\frac{D_V}{r_d})}+\chi^2_{(\frac{D_H}{r_d})}]$.\\

\renewcommand{\arraystretch}{1.0}
\begin{table}[H]
    \centering
    \begin{tabular}{|c|c|c|c|c|}
    \hline
       $z$ & \textbf{$H(z)$} & \textbf{$\sigma_{H,i}$} & \textbf{Method} & \textbf{Refs.} \\
        \hline
        0.07 & 69.0 & $\pm 19.60$ & CC &  \cite{zhang}\\
        \hline
         0.0708 & 69.0 & $\pm 19.68$ & CC & \cite{Abbott1}\\
        \hline
         0.09 & 69.0 & $\pm 12.0$ & CC & \cite{Jimenez}\\
        \hline
         0.12 & 68.6 & $\pm 26.2$ & CC &  \cite{zhang}\\
        \hline
         0.17 & 83.0 & $\pm 8.00$& CC &  \cite{Simon}\\
        \hline
         0.179 & 75.0 & $\pm 4.0$ & CC & \cite{Moresco}\\
        \hline
         0.199 & 75.0 & $\pm 5.0$ & CC & \cite{Moresco}\\
        \hline
         0.20 & 72.9 & $\pm 29.6$ & CC & \cite{Abbott1}\\
        \hline
         0.24 & 79.69 & $\pm 3.32$ & BAO &  \cite{Hui}\\
        \hline
         0.27 & 77.0 & $\pm 14.0$ & CC &  \cite{Simon}\\
        \hline
         0.28 & 88.8 & $\pm 36.6$ & CC & \cite{Abbott1}\\
        \hline
         0.35 & 84.4 & $\pm 7.0$ & BAO &  \cite{Xu}\\
        \hline
         0.352 & 83.0 & $\pm 14.0$ & CC &  \cite{Moresco}\\
        \hline
         0.38 & 81.2 & $\pm 2.2$ & CC & \cite{Metin}\\
        \hline
         0.3802 & 83.0 & $\pm 14.0$ & CC & \cite{Eric}\\
        \hline
         0.4 & 95.0 & $\pm 17.0$ & CC &  \cite{Simon}\\
        \hline
         0.4004 & 77.0 & $\pm 10.2$ & CC & \cite{Eric}\\
        \hline
         0.4247 & 87.1 & $\pm 11.2$ & CC & \cite{Eric}\\
        \hline
         0.43 & 86.45 & $\pm 3.27$ & BAO &  \cite{Hui}\\
        \hline
         0.44 & 82.6 & $\pm 7.8$ & BAO &   \cite{Blake}\\
        \hline
         0.4497 & 92.8 & $\pm 12.9$ & CC & \cite{Eric}\\
        \hline
         0.47 & 89.0 & $\pm 50.0$ & CC & \cite{Julian}\\
        \hline
         0.4783 & 80.9 & $\pm 9.0$ & CC &  \cite{Eric}\\
        \hline
         0.48 & 97.0 & $\pm 62.0$ & CC &  \cite{Stern}\\
        \hline
         0.51 & 90.9 & $\pm 2.1$ & CC & \cite{Metin}\\
        \hline
         0.57 & 92.4 & $\pm 4.5$ & BAO & \cite{Samushia}\\
        \hline
         0.593 & 104.0 & $\pm 13.0$ & CC & \cite{Moresco}\\
        \hline
         0.6 & 87.9 & $\pm 6.1$ & BAO &  \cite{Blake}\\
        \hline
         0.61 & 98.96 & $\pm 2.20$ & CC & \cite{Metin}\\
        \hline
         0.68 & 92.00 & $\pm 8.0$ & CC &  \cite{Moresco}\\
        \hline
         0.73 & 97.3 & $\pm 7.0$ & BAO &  \cite{Blake}\\
        \hline
         0.781 & 105.0 & $\pm 12.0$ & CC &  \cite{Moresco}\\
        \hline
         0.875 & 125.0 & $\pm 17.0$ & CC &  \cite{Moresco}\\
        \hline
         0.88 & 99.0 & $\pm 40.0$ & CC &  \cite{Stern}\\
        \hline
         0.9 & 117.0 & $\pm 23.0$ & CC &  \cite{Simon}\\
        \hline
         1.037 & 154.0 & $\pm 20.0$ & CC & \cite{Moresco}\\
        \hline
         1.3 & 168.0 & $\pm 17.0$ & CC & \cite{Simon}\\
        \hline
         1.363 & 160.0 & $\pm 33.6$ & CC & \cite{Betoule}\\
        \hline
         1.43 & 177.0 & $\pm 18.0$ & CC &  \cite{Simon}\\
        \hline
         1.53 & 140.0 & $\pm 14.0$ & CC &  \cite{Simon}\\
        \hline
         1.75 & 202.0 & $\pm 40.0$ & CC &  \cite{Simon}\\
        \hline
         1.965 & 186.5 & $\pm 50.4$ & CC & \cite{Betoule}\\
        \hline
         2.34 & 222.0 & $\pm 7.0$ & BAO & \cite{Beutler1}\\
        \hline
         2.36 & 226.0 & $\pm 8.0$ & BAO & \cite{Beutler2}\\
         \hline
    \end{tabular}
    \caption{Here the unit of $H(z)$ is $km.s^{-1}.Mpc^{-1}$, 'CC' indicates that the $H(z)$ value is derived from the cosmic chronological method or the differential age approach, while 'BAO' signifies that the $H(z)$ value is acquired using Baryon Acoustic Oscillations data, with the relevant references cited accordingly.}
    \label{tab:I}
\end{table}

\renewcommand{\arraystretch}{1.2}
\begin{table*}[t]
    \centering
    \begin{tabular}{|c|c|c|c|c|c|c|c|c|}
    \hline
        \textbf{$\Tilde{z}$} & \textbf{0.15} & \textbf{0.38} & \textbf{0.51} & \textbf{0.70} & \textbf{0.85} & \textbf{1.48} & \textbf{2.33} & \textbf{2.33} \\
        \hline
       $D_V(\Tilde{z})/r_d$ & $4.47\pm 0.17$ & - & - & -&$18.33^{+0.57}_{-0.62}$ & - & - & - \\
        \hline
      $D_M(\Tilde{z})/r_d$ & - & $10.23\pm 0.17$ & $13.36\pm 0.21$ & $17.86\pm 0.33$ & - & $30.69\pm 0.80$ & $37.6\pm 1.9$ & $37.3\pm 1.7$ \\
        \hline
        $D_H(\Tilde{z})/r_d$ & - & $25.00\pm 0.76$ & $22.33\pm 0.58$ & $19.33\pm 0.53$ & - & $13.26\pm 0.55$ & $8.93\pm 0.28$ & $9.08\pm 0.34$ \\
        \hline
    \end{tabular}
    \caption{SDSSBAO Dataset}
    \label{tab:II}
\end{table*}    

\renewcommand{\arraystretch}{1.2}
\begin{table}[t]
    \centering
    \begin{tabular}{|c|c|c|c|}
    \hline
        \textbf{$\Tilde{z}$} & \textbf{ $D_V(\Tilde{z})/r_d$} & \textbf{ $D_M(\Tilde{z})/r_d$} & \textbf{$D_H(\Tilde{z})/r_d$} \\
        \hline
        $0.295$ & $7.942\pm 0.075$ &- &- \\
        \hline
        $0.510$ & $12.720\pm 0.099$ & $13.588\pm 0.167$ & $21.863\pm 0.425$ \\
        \hline
        $0.706$ & $16.050\pm 0.110$ & $17.351\pm 0.177$ & $19.455\pm 0.330$ \\
        \hline
        $0.922$ & $19.656\pm 0.105$ & $21.648\pm 0.178$ & $17.577\pm 0.213$ \\
        \hline
        $0.934$ & $19.721\pm 0.091$ & $21.576\pm 0.152$ & $17.641\pm 0.193$ \\
        \hline
        $0.955$ & $20.008\pm 0.183$ & $21.707\pm 0.335$ & $17.803\pm 0.297$ \\
        \hline
        $1.321$ & $24.252\pm 0.174$ & $27.601\pm 0.318$ & $14.176\pm 0.221$ \\
        \hline
        $1.484$ & $26.055\pm 0.398$ & $30.512\pm 0.760$ & $12.817\pm 0.516$ \\
        \hline
        $2.330$ & $31.267\pm 0.256$ & $38.988\pm 0.531$ & $8.632\pm 0.101$ \\
        \hline
    \end{tabular}
    \caption{DESBAO Dataset (DR2)}
    \label{tab:III}
\end{table}

\begin{figure}
\centering
\includegraphics[width=8cm, height=4.5cm]{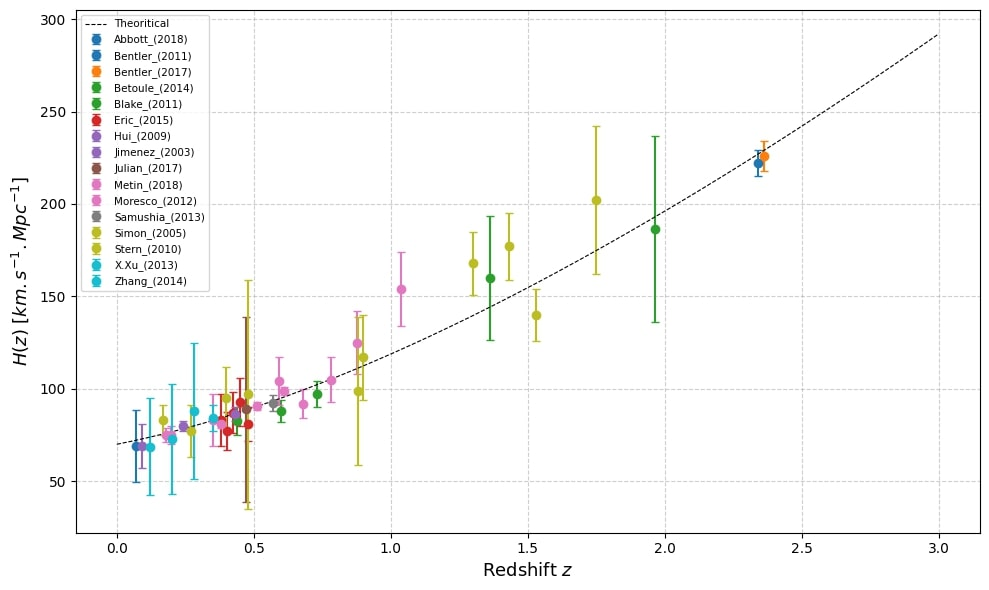}
 \caption{All available OHD points are in Table \ref{tab:I}. The dashed line plots the theoretical Hubble parameter $H(z)$ as a function of $z$ from the spatially flat $\Lambda$CDM model with typical values of  $\Omega_{DM,0}=0.3$, $\Omega_{DE,0}=0.7$ and $H_0 =$70 $km. s^{-1} Mpc^{-1}$.}\label{Fig1} 
\end{figure}

The plot of the Hubble parameter $H(z)$ as a function of redshift $z$ in Fig. \ref{Fig1}, incorporating both the theoretical prediction curve and OHD with associated uncertainties, serves as a critical diagnostic tool for evaluating cosmological models. {\it Although our investigation via the alternate GW-inspired RE in Eq. (\ref{31}) is distinct from the standard Friedmann formalism via the usual $\Lambda$CDM cosmology, the theoretical curve in that plot successfully reproduces the observed expansion history.} Therefore, Fig. \ref{Fig1} indicates that the GW-inspired RE Eq. (\ref{20}), when formulated in redshift space, is fully capable of capturing the dynamical behavior of the expansion of the universe and illustrates that our model fits accurately with the Hubble measurement \cite{Bouali}. The inclusion of error bars in the OHD points captures the observational uncertainties ($\sigma_{H,i}$), allowing for a statistically meaningful comparison. This visual and quantitative analysis facilitates constraints on model parameters and aids in probing the underlying dynamics of dark energy and cosmic acceleration. Thus, the agreement between our model and observational data highlights the consistency and predictive ability of Eq. (\ref{31}) in cosmological modeling through Fig. \ref{Fig1}.   

We used Bayesian inference with the No-U-Turn Sampler (NUTS) \cite{Gelman} to estimate cosmological parameters ($p$) from OHD. Typically, the Markov Chain Monte Carlo (MCMC) \cite{Lewis} sampling produced posterior probability distributions for each model parameter, enabling a thorough assessment of parameter uncertainty within a Bayesian statistical framework. The best fit values of the model parameters were ascertained from the posterior mean estimations obtained using MCMC sampling. The model's goodness-of-fit was quantitatively evaluated using chi-squared ($\chi^2$) statistics, which compare the model predictions to the observed data. Additionally, the marginalized posterior distributions were shown using contour plots, highlighting the $1\sigma$ (68\%) and $2\sigma$ (95\%) credible regions to depict the confidence intervals and parameter degeneracies inside the multidimensional parameter space.

\begin{figure}[H]
\centering
\includegraphics[width=7cm, height=7cm]{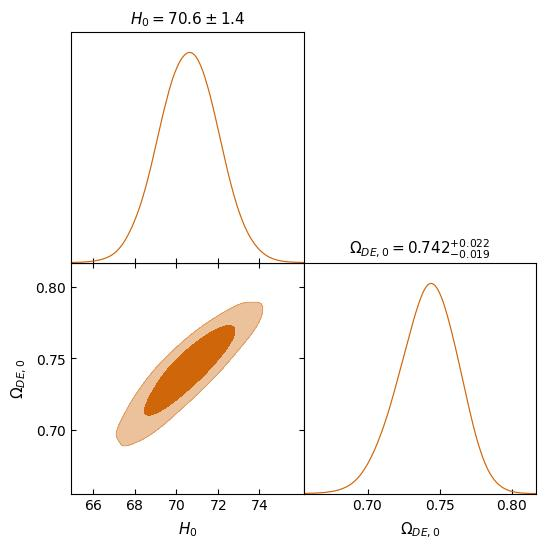}
\caption{Fitting of model parameters to Hubble data.}\label{Fig2}
\end{figure}

\begin{figure}
\centering
\includegraphics[width=7cm, height=7cm]{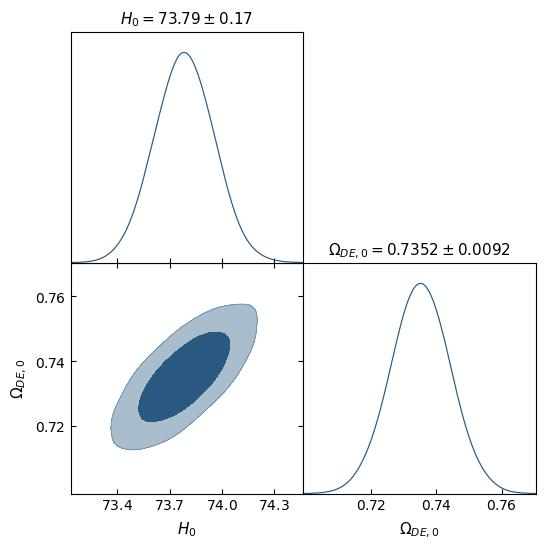}
\caption{Fitting of model parameters to Pantheon+Hubble data.}\label{Fig3}
\end{figure}

\begin{figure}
\centering
\includegraphics[width=8cm, height=8cm]{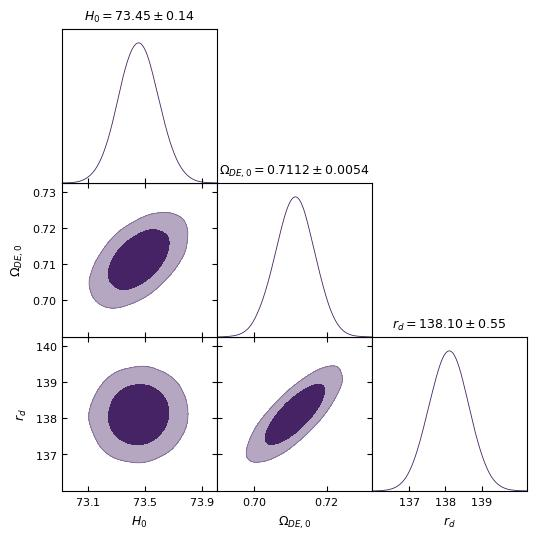}
\caption{Fitting of model parameters to Pantheon+Hubble+BAO data.}\label{Fig4}
\end{figure}

The initial conditions or prior distribution of parameters are presented in Table \ref{tab:IV}. Here, `Uniform' represents a uniform distribution in the interval (lower and upper limits). According to observations of the cosmic microwave background (CMB), assuming a flat $\Lambda$CDM cosmology, yields a Hubble constant of $H_0= 67.36\pm 0.54\ \mathrm{km.s^{-1}.Mpc^{-1}}$ as reported by Planck Collaboration 2018 \cite{Planck1, Planck2}. This result was refined in the Planck 2018 Legacy Release \cite{Planck3} (publicly made available in 2020), which continues to support a consistent value within uncertainties: $H_0 \approx 67.4\pm 0.5\ \mathrm{km.s^{-1}.Mpc^{-1}}$, reinforcing the tension with the late-time measurements such as those from SHOES data: $H_0\approx73.2\pm 1.3\ \mathrm{km.s^{-1}.Mpc^{-1}}$ \cite{Riess}.

\renewcommand{\arraystretch}{1.2}
\begin{table*}[t]
    \centering
    \begin{tabular}{|c|c|c|c|c|c|}
    \hline
        \textbf{Parameters} & \textbf{Prior Distribution} & \textbf{Hubble data} & \textbf{Pantheon+Hubble data} & \textbf{Pantheon+Hubble+BAO data} \\
        \hline 
         $H_0$ & Uniform(50, 100) & $70.6\pm 1.4$ & $73.79\pm 0.17$ & $73.45\pm 0.14$ \\
        \hline
        $\Omega_{DE,0}$ & Uniform(0.5,0.9) & $0.742^{+0.022}_{-0.019}$ &$0.7352\pm 0.0092$ & $0.7112\pm0.0054$ \\
        \hline
        $r_d$ & Uniform(100,300) & - & - & $138.10\pm0.55$\\  
        \hline
    \end{tabular}
    \caption{List of cosmological parameters with there initial conditions and obtained best-fit values.}
    \label{tab:IV}
\end{table*}

To constrain the cosmological parameters, we performed a Bayesian parameter estimation using the MCMC technique. Our cosmological parameter estimation highlights the power of combining multiple observational datasets to yield progressively tighter constraints on key parameters governing the universe's expansion. As summarized in Table \ref{tab:IV}, we analyze three primary parameters $H_0$, $\Omega_{DE,0}$, and $r_d$, using various combinations of Hubble, {\it{Pantheon}} and BAO datasets under a flat $\Lambda$CDM cosmological model. Using only OHD, Fig. \ref{Fig2} results in the lower value,  $H_0=70.6\pm 1.4$ $\mathrm{km.s^{-1}.Mpc^{-1}}$, which reflects the direct measurements of cosmic expansion. The Fig. \ref{Fig3} is constructed using the {\it Pantheon+Hubble} combined dataset and obtains a comparatively higher constraint of $H_0=73.79\pm 0.17$ $\mathrm{km.s^{-1}.Mpc^{-1}}$. Then Fig. \ref{Fig4}, which includes the {\it Hubble, Pantheon, and BAO} data together, yields the tightest constraint with $H_0=73.45\pm 0.14$ $\mathrm{km.s^{-1}.Mpc^{-1}}$, which indicates a significantly high late-time expansion rate of the universe. This value exemplifies the discrepancy between late-universe measurements and early-universe inferences of the Hubble constant. In particular, it stands in strong tension with the value inferred from CMB observed by Planck 2018 \cite{Planck2}, which favors a much lower value of $H_0= 67.4\pm 0.5$ $\mathrm{km.s^{-1}.Mpc^{-1}}$ . This long-standing discrepancy, known as the Hubble Tension, persists despite improvements in late-time and CMB-based measurements. Its statistical significance and robustness indicate the need for extension to the standard cosmological model, such as {\it{Early Dark Energy (EDE)}}, which could increase the expansion rate before recombination,  {\it{Interacting Dark Sectors}}, modified gravity, or other beyond-$\Lambda$CDM physics.   

Mainly the corner plot in Fig. \ref{Fig4}, additionally predicts sound horizon at the drag epoch, $r_d= 138.10\pm 0.55$ Mpc. Here, the higher inference value of $H_0=73.45\pm 0.14$ $\mathrm{km.s^{-1}.Mpc^{-1}}$ is notably consistent with local universe measurements such as those from the SHOES collaboration \cite{Riess}. Simultaneously, the lower value of $r_d~(=138.10\pm 0.55)$, closely aligning with early-universe expansion history as predicted by $\Lambda$CDM model ($139.5^{+5.2}_{-4.4}$ Mpc \cite{Liu}) and ($r_d \approx 137.44\pm 0.64$ Mpc \cite{Ganguly1}), alongside accelerated late time expansion. This compensatory mechanism—higher $H_0$ and shorter $r_d$ is a critical clue pointing towards new physics beyond the standard $\Lambda$CDM model.   

Here we have also investigated the variation of the deceleration parameter $q(z)$ concerning redshift $z$. This is a dimensionless quantity that describes the rate at which the expansion of the universe is slowing down or speeding up. A positive $q(z)$ indicates a decelerating universe, while a negative $q(z)$ indicates an accelerating universe. This deceleration parameter can be expressed:
\ben
&&q= -\frac{1}{H^2} \Big(\frac{\ddot{a}a}{a}\Big)=-\Big(\frac{\ddot{a}a}{\dot{a}^2} \Big) \nonumber\\ && \Rightarrow q(z)=-1+\frac{(1+z)}{H(z)}\Big(\frac{dH}{dz}\Big) \nonumber\\
&&\Rightarrow q(z)=\frac{1}{2} [1-3 \Omega_{DE,0}] \label{45}
\een

Now we have plotted $q(z)$ {\it vs.} $z$ by using the best-fit parameter obtained from {\it Pantheon+ Hubble+ BAO} combined data set \cite{Alam, Eisenstein1, Percival, Glazebrook, Hu, Eisenstein2, Dawson, DESI}.

\begin{figure}[H]
\centering
\includegraphics[width=9cm, height=5.5cm]{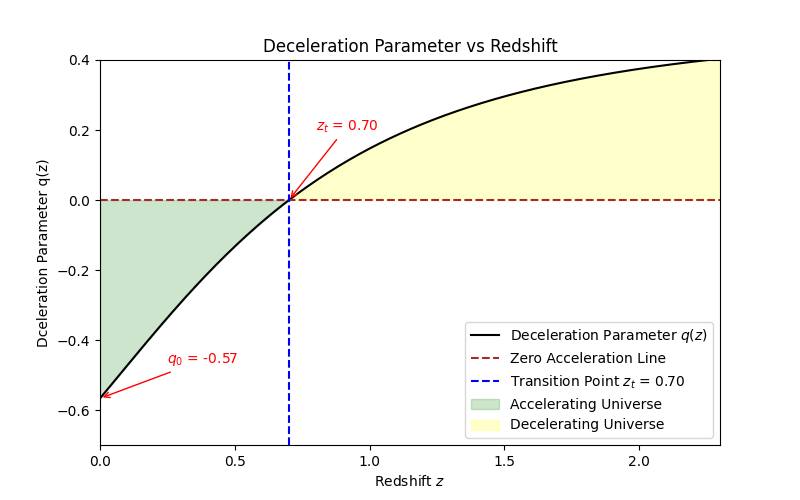}
\caption{Deceleration parameter $q(z)$ vs. Redshift ($z$) plot with parameters $H_0\approx73.45$ and $\Omega_{DE,0}\approx0.7112$.}\label{Fig5}
\end{figure}

From Fig. \ref{Fig5} we can tell that, at high redshifts, especially for $z>0.70$, the cosmic expansion undergoes a deceleration phase, as quantified by the positive deceleration parameter $q>0$. This deceleration is a direct consequence of the attractive gravitational influence of matter, which acts to slow down the expansion rate. This matter-dominated regime is clearly illustrated in the yellow-shaded region of Fig. (\ref{Fig5}), representing the epoch prior to the onset of the accelerated expansion. Now the transition redshift, $z_t\approx 0.70$, marking a critical epoch in the cosmic history where the contributions from the dark energy become dynamically significant. At this juncture, the negative pressure component associated with dark energy begins to counterbalance the gravitational pull of matter. The deceleration parameter approaches zero, signaling the shift from a phase of decelerating to a phase of accelerating universe. For redshift $z<0.70$, the cosmological constant (or an equivalent dark energy component with an equation of state close to $\omega=-1$) dominates the energy budget, driving a phase of accelerated expansion. This acceleration is characterized by a negative deceleration parameter $q<0$, as indicated by the green-shaded region in Fig. (\ref{Fig5}. The present-day value of $q(z)$ i.e. $q_0=-0.57$, derived from our model fits, corroborated a universe undergoing acceleration, in excellent agreement with the concordance $\Lambda$CDM paradigm ($q_0\approx -0.5$ \cite{Weinberg1, Planck3}).

We have reexamined the reformulation of the modified GW-inspired RE Eq. (\ref{22}) into a harmonic oscillator model \cite{Kar, Dadhich, Ganguly}. This modeling approach may provide significant scientific insights about the impact of RE on cosmic evolution, depicting time-like geodesic congruence. It should be noted that the expansion $\Theta$ represents the rate of change of the cross-sectional area orthogonal to the bundle of geodesics. The expansion approaching negative infinity indicates the bundle's convergence, whereas a positive infinity value signifies total divergence. 

At first, we redefine $\Theta=3\frac{\dot{f}}{f}$, where $f$ is the function of the cosmic time $t$, related to the expansion of a bundle of flow curves. So, using the above definition of $\Theta$, we can write the following expression from Eqs. (\ref{22}) and (\ref{24}),
\ben
\Big(\frac{d^2f}{dt^2}\Big)+ \Big[\frac{4\pi G}{3}(\rho+3P)\Big]f=0 \label{46}
\een
This Eq. (\ref{46}) represents a Hill-type equation, or a harmonic oscillator equation, facilitating the analysis of expansion theory. Here, we assume the universe is homogeneous and isotropic, resulting in a zero rotation tensor $\omega_{\alpha\beta}$ (as previously assumed). Consequently, the inference of convergence simplifies for {\it{hypersurface orthogonal congruences}} ($R_{\alpha\beta} u^\alpha u^\beta\geq 0$), which leads to {\it geodesic focusing}. Detailed explanations can be found in \cite{Kar, Dadhich, Ganguly}.

In the harmonic oscillator Eq. (\ref{46}), we may denote $\omega^2(t)=\Big[\frac{4 \pi G}{3} (\rho+3P)\Big]$, where $\omega^2(t)$ denotes a time-dependent frequency, signifying the dynamical impact of spacetime. Now, if $\omega^2(t)\geq0$, which signifies the Strong Energy Condition (SEC) \cite{Kar}, indicating that oscillation occurs under attractive gravity and that the expansion of the universe is decelerating. If $\omega^2(t)$ is negative, repulsive effects prevail, potentially resulting in exponential divergence, as seen in inflation, anti-trapped surfaces (inside black hole interiors), or violations of energy conditions \cite{Ganguly}. The expansion of the cosmos is accelerating.

Now we can express the Eq. (\ref{46}) in terms of the redshift parameter $z$ by using the relation from Eq. (\ref{28}) as the following form
\ben 
\Big(\frac{d^2f}{dz^2}\Big) &+&\Big[\frac{2}{(1+z)}+ \frac{(1-3\Omega_{DE})}{2(1+z)}\Big]\Big(\frac{df}{dz}\Big)\nonumber\\&+&\Big[\frac{(1-3\Omega_{DE})}{2(1+z)}\Big]f=0 \label{47}
\een 
This Eq. (\ref{47}) is mathematically analogous to the damped harmonic oscillator equation. Here the coefficient of $\Big(\frac{df}{dz}\Big)$ acts like a damping term, depending on $z$ and $\Omega_{DE}$. 

\begin{figure}[H]
\centering
\includegraphics[width=8cm, height=5.5cm]{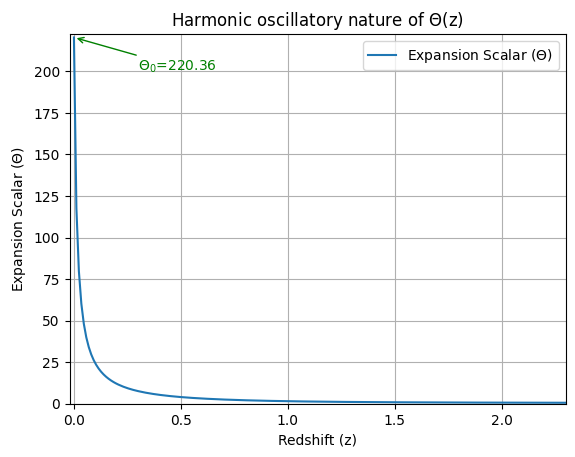}
\caption{Expansion scalar($\Theta$) $vs.$ Redshift ($z$) plot with parameters $H_0=73.45$.}\label{Fig6}
\end{figure}

We solve this equation (\ref{47}) numerically and plot $\Theta(z)$ $vs.$ $z$ in Fig. \ref{Fig6} where we put the present value of the expansion scalar $\Theta_0=220.36$ corresponding to the best fit parameter $H_0=73.45$ $km.s^{-1}.Mpc^{-1}$ obtained from {\it the Pantheon+Hubble+BAO} dataset. This plot in Fig. \ref{Fig6} represents the rapid decline of $\Theta(z)$ with increasing redshift value, which implies that, in the early universe, the expansion was slower in terms of volume expansion scalar, but towards the present time ($z=0$) accelerated expansion causes $\Theta$ to increase, consistent with a dark-energy-dominated late universe.\\

{\textbf{Observational Data Analysis with Nuisance Parameter $\mu_0$}}:
For enhancing the accuracy and flexibility of cosmological parameter estimation from Type Ia supernova (SNe Ia) data, we introduce a nuisance parameter $\mu_0$, which encapsulates the combined degeneracy between the absolute magnitude $M$ of SNe Ia and $H_0$. Instead of assuming a fixed offset, we incorporate these uncertainties into a single additive parameter $\mu_0$, which is treated as a free parameter and marginalized over during the inference process. This $\mu_0$ is called a nuisance parameter. Specifically, we have modified Eq. (\ref{35}) into the following form \cite{Ganguly1}:
\ben
\mu(z,\theta)= 5~log_{10}(d_L(z))+ \mu_0 \label{48}
\een
where $\mu_0= 5~log_{10}(\frac{c}{H_0})+25+M$, $c$ is the speed of light in km/s, and $M$ accounts for the absolute magnitude calibration of supernovae. By using $\mu_0$ as a free parameter in the Bayesian process, we have achieved a model-independent fit that more faithfully reflects the structure of the data. 

\begin{figure}[H]
\centering
\includegraphics[width=8cm, height=8cm]{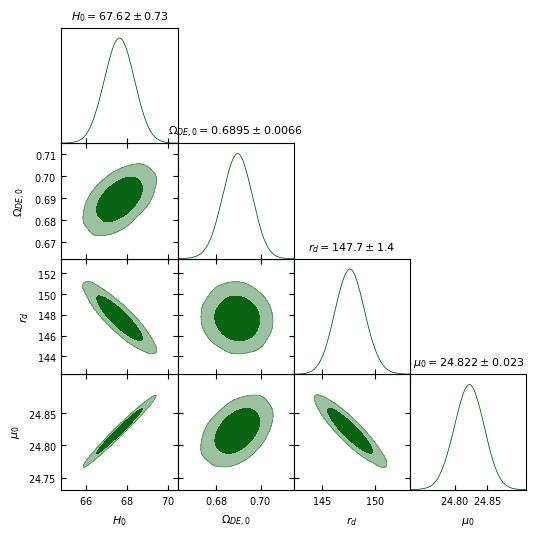}
\caption{Fitting of model parameters to Pantheon+Hubble+BAO data with nuisence parameter $\mu_0$.}\label{Fig7}
\end{figure}

Fig. {\ref{Fig7}} illustrates the Bayesian analysis of our model inside a four-parameter structure: $H_0,~\Omega_{DE,0},~r_d$, and $\mu_0$. This corner plot demonstrates well-defined posterior distributions with minimal uncertainty and almost Gaussian marginal contours, signifying a solid statistical fit to the observational data. The inclusion of the nuisance parameter $\mu_0$ results in a slight broadening of the marginalization distributions, indicating its function in accounting for systematic uncertainties related to absolute magnitude calibration and the Hubble constant. Despite this, the model maintains strong internal consistency and demonstrates a high degree of concordance with current cosmological observations, reinforcing its validity within the assumed parameter space.  

\renewcommand{\arraystretch}{1.5}
\begin{table}[h]
    \centering
    \begin{tabular}{|c|c|c|}
    \hline
        \textbf{Parameters} & \textbf{Prior Distribution} & \textbf{Obtained Values} \\
         \hline 
         $H_0$ & Uniform(50, 100) & $67.62\pm 0.73$  \\
        \hline
        $\Omega_{DE,0}$ & Uniform(0.5,0.9) & $0.6895\pm 0.0066$  \\
        \hline
        $r_d$ & Uniform(100,300) & $147.7\pm1.4$  \\
        \hline
        $\mu_0$ & Uniform(20,30) &  $24.822\pm0.023$\\  
        \hline
       
    \end{tabular}
    \caption{List of cosmological parameters, including additional nuisance term $\mu_0$, with four initial conditions and obtained best-fit values.}
    \label{tab:V}
\end{table}

Here, Table \ref{tab:V} summarizes the best fit values obtained from Fig. \ref{Fig7} for the model parameters $H_0$, $\Omega_{DE,0}$, and $r_d$, along with the nuisance parameter $\mu_0$, using uniform prior distributions. The estimated values mentioned in that table reflect a statistically well-constrained inference, indicating a strong internal consistency when accounting for calibration uncertainties through $\mu_0$.

\begin{figure}[H]
\centering
\includegraphics[width=9cm, height=5.5cm]{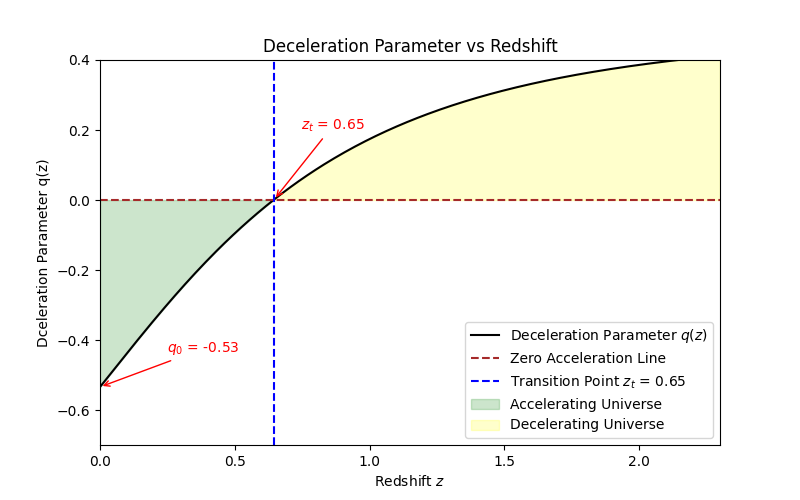}
\caption{Deceleration parameter $q(z)$ vs. Redshift ($z$) plot with parameters $H_0\approx67.62$ and $\Omega_{DE,0}\approx0.6895$ and $\mu_0\approx24.822$}\label{Fig8}
\end{figure}

Here again we have examined the rate of change of the deceleration parameter $q(z)$ as a function of redshift $z$ in Fig. \ref{Fig8}, but this time we consider the additive nuisance parameter $\mu_0$. Through this, we get the new value of transition redshift, $z_t \approx 0.65$, and at present $q(z)$, i.e., $q_0=-0.53$.

To assess the cosmological viability of the deceleration parameter models under consideration, we compare the inferred $q_0$, and $z_t$ from our plots, Fig. \ref{Fig5} and Fig. \ref{Fig8}. The values of $q_0$ and $z_t$, mentioned earlier, are broadly consistent with constraints derived from recent observational studies. Specifically, Capozziello et al. (2014) \cite{Capozziello}  reported $z_t\approx 0.69-0.75$. Furthermore, in the standard flat $\Lambda$CDM model, $q_0\approx -0.55$ in agreement with Planck 2018 \cite{Planck1} results. So here, both our considerations (Figs. \ref{Fig5} and \ref{Fig7}) thus produce values of $q_0$ and $z_t$ that fall within the observationally allowed ranges, indicating their consistency with the current cosmological paradigm.  

\renewcommand{\arraystretch}{1.5}
\begin{table}[htbp]
\centering
\begin{tabular}{|c|c|c|c|c|}
\hline
\textbf{Parameters} & \multicolumn{2}{c|}{\textbf{From Table IV}} & \multicolumn{2}{c|}{\textbf{From Table V}} \\
\cline{2-5}
 & \textbf{Mean} & \textbf{Std. Dev.} & \textbf{Mean} & \textbf{Std. Dev.} \\
\hline
$H_0$ & 73.45 & $\pm\,0.14$ & 67.62 & $\pm\,0.73$ \\
\hline
$\Omega_{{DE},0}$ & 0.7112 & $\pm\,0.0054$ & 0.6895 & $\pm\,0.0066$ \\
\hline
$r_d$ & 138.10 & $\pm\,0.55$ & 147.67 & $\pm\,1.41$ \\
\hline
$\mu_0$ & -- & -- & 24.822 & $\pm\,0.023$ \\
\hline
\end{tabular}
\caption{Side-by-side comparison of parameters estimated using best-fit {\it Pantheon+Hubble+BAO} data without and with  nuisance parameter $\mu_0$, collected from Tables \ref{tab:IV} and \ref{tab:V}, respectively.}
\label{tab:VI}
\end{table}

Table \ref{tab:VI} provides a side-by-side comparison of parameters estimated using best-fit {\it Pantheon+Hubble+BAO} data, showing that, without considering $\mu_0$, the estimated Hubble constant $H_0$ is significantly higher at $73.45\pm 0.14$  $km.s^{-1}.Mpc^{-1}$ and  $\Omega_{DE,0}=0.7112\pm0.0054$, both with much smaller uncertainties, are consistent with the SHOES measurements \cite{Riess} and $r_d=138.10\pm0.55$ align with the results predicted by \cite{Ganguly, Ganguly1}. However, the inclusion of $\mu_0$ reduces $H_0=67.62\pm 0.73$ $km.s^{-1}.Mpc^{-1}$ and $\Omega_{DE,0}=0.6895 \pm 0.0066$, the values to be more consistent with CMB-based measurements (e.g. Planck 2018 \cite{Planck1}), suggesting a potential indication of reconciliation of the Hubble tension. In addition to this, the value of $r_d=147.7\pm1.4$ is very close to that predicted by $\Lambda$CDM model \cite{Planck2}. Additionally, while uncertainties in parameters like $H_0$ and $r_d$ increase with $\mu_0$, this broadening accounts for systemic effects, making the model statistically more flexible.

Notably, the inclusion of $\mu_0$ significantly reduces the estimated value of $H_0$, bringing it closer to early universe observations from the CMB \cite{Planck1}. The analysis without $\mu_0$ produces a $H_0$ value that is more closely matched with local distance ladder measurements \cite{Riess2}. This variation is physically relevant because it illustrates the underlying conflict between early and late-time measurements of $H_0$ and demonstrates how differing model assumptions might alter cosmological inference. This tendency indicates the base of the Hubble tension, which is the dependency of the $H_0$ estimates on calibration assumptions and observational systematics. Essentially, the nuisance parameter $\mu_0$ serves as a buffer for these uncertainties. As $H_0$ and $M$ are degenerate in the distance modulus, $\mu_0$ effectively absorbs this degeneracy, allowing for more accurate determination of cosmological parameters, particularly $H_0$. By marginalizing over $\mu_0$, we decoupled model-dependent assumptions from observational systematics, improving the physical interpretability of the inferred cosmic constraints.

Physically, $\mu_0$ acts as a normalizing factor for the luminosity distance, without affecting the modified GW-inspired RE dynamics. Its purpose is essentially phenomenological, absorbing systematic offsets in the distance-redshift connection caused by calibration errors or dataset-dependent normalization discrepancies. However, $\mu_0$ has some restrictions. It is primarily a calibration parameter, used to ignore zero-point offsets in the distance modulus relation and misinterpretability of shifts in $\mu_0$ across datasets, rather than a resolution of the Hubble tension \cite{Roberto}.

Overall, the inclusion of $\mu_0$ introduces a more flexible modeling framework that better encapsulates observational uncertainties, particularly in supernova luminosity calibration. This leads to parameter estimates more consistent with early-universe probes, albeit with slightly larger posterior variances. \\ 

\begin{figure}[H]
\centering
\includegraphics[width=8cm, height=5.5cm]{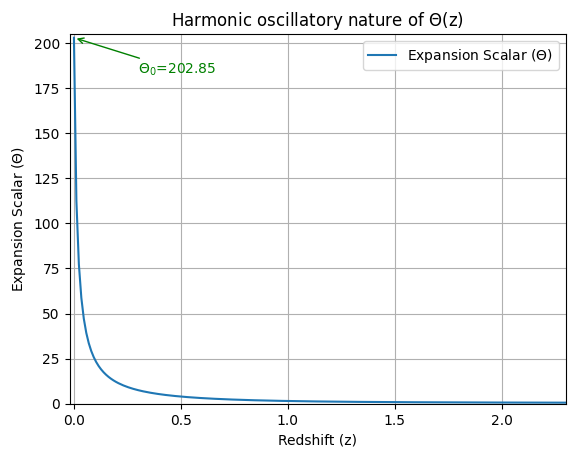}
\caption{Expansion scalar($\Theta$) $vs.$ Redshift ($z$) plot with parameters $H_0=67.62$ .}\label{Fig9}
\end{figure}

Now we have also plotted the graph $\Theta(z)$ $vs.$ $z$  in Fig. \ref{Fig9} considering the nuisance parameter, using the Eq. (\ref{47}) as before, where $\Theta_0=202.85$ is calculated from the best-fit parameter $H_0=67.62$ $km.s^{-1}.Mpc^{-1}$, mentioned in Table \ref{tab:V}, obtained from {\it Pantheon+ Hubble+ BAO data}. Similarly as Fig. \ref{Fig6}, this plot in Fig. \ref{Fig9} can explain that the drastic increasing rate of $\Theta(z)$ with decreasing redshift $z$ reflects the transition from the decelerated expansion regime in the matter-dominated early universe to the accelerated phase governed by the dark energy-dominated late universe, which concept has been explained in Figs. \ref{Fig5} and \ref{Fig8}. These formulations connect kinematical quantities like the expansion scalar to cosmological dynamics via GW-inspired RE.

Beyond its pedagogical and heuristic value, the reinterpretation of the GW-inspired RE as a harmonic oscillator ((\ref{46}) or (\ref{47})) can be contextualized within the broader literature on possible resolution of Hubble tension. 

For example, Early Dark Energy (EDE) models propose the existence of an extra transient dark energy term that exists only for a short period of the early universe ($z \gtrsim 3000$). During this brief epoch, the additional energy density increases the early-universe expansion rate and reduces the sound horizon $r_d$ at the recombination, thereby allowing the CMB-inferred value of $H_0$ to agree more closely with late-time measurements \cite{Paulin}. Side by side we can mention an important variation of EDE that is the Chameleon Early Dark Energy model (CEDE) \cite{Karwal} which embeds the EDE in a Chameleon-like scaler field potential, where Chameleon field is a type of scaler field whose effective mass depends on the local matter density, allowing the field to interact with matter in a way, that hides its effects in high density regions but still changes the cosmic expansion rate in the early universe. In contrast, our GW-inspired RE operates within standard GR but incorporates gravitational wave–driven modifications to the kinematics via small–metric–perturbation corrections ($h_{\mu\nu}$), producing similar phenomenological signatures (higher late–time $H_0$ with smaller $r_d$) without introducing new fundamental degrees of freedom. 

On the other hand, Interacting Dark Sector models assume that dark matter and dark energy are not completely separated, but this model posits energy-momentum exchange between dark matter and dark energy.  This is done by adding a coupling term to their continuity equation. The strength and the form of this coupling determine how the energy flows between these two components. Thus, it changes the expansion history $H(z)$, especially at late times, and can raise the value of $H_0$ inferred from observations \cite{Ganguly1, Valentino, Pan, He}. However, these models require choosing a specific mathematical form for the coupling and assume a non-gravitational interaction between dark matter and dark energy. But our GW-inspired RE approach does not introduce any new couplings or forces. The changes in the expansion rate arise intrinsically from the GW perturbation term in the RE, highlighting geometric effects, meaning the mechanism stays entirely within the standard gravitational framework.\\

Now we are going to discuss a statistical comparison to judge which of the methods (with and without nuisance parameter $\mu_0$) best fits the observational data. Based on our observational data analysis, we have introduced different statistical criteria such as, $\chi^2$, the total chi-squared value and $\chi^2_\nu$, the reduced chi-squared value (defined as $\chi^2_\nu= \chi^2/\nu$), where $\nu$ is the number of degrees of freedom, which can be defined as $\nu=N-k$, where $N$ is the total number of data points and $k$ is the number of free model parameters. We have two different types of information criteria, such as the Akaike Information Criterion (AIC) and the Bayesian Information Criterion (BIC). They can also be defined as AIC $(=\chi^2+2k)$ and BIC $(=\chi^2+k \ln N)$. 

In our case, the number of data points ($N$) for both the models is the same ($N=1783$), while the number of free parameters is $k=4$ and $k=3$ for the different modeling methods, with and without $\mu_0$, respectively. To evaluate and compare the performance of these two methods, we have analyzed their statistical metrics: $\chi^2$, $\chi^2_\nu$, AIC, and BIC, which are summarized in Table \ref{tab:VII}. These criteria measure how well the model fits the data while also penalizing models that are too complex. This helps in selecting a model that explains the data well without using too many parameters, avoiding overfitting.

\renewcommand{\arraystretch}{1.5}
\begin{table}[h]
    \centering
    \begin{tabular}{|c|c|c|c|c|}
    \hline
        \textbf{Method} & $\chi^2$ & $\chi^2_\nu$ & AIC & BIC \\
         \hline 
         Without $\mu_0$ & $1895.3365$ & $1.0648$ & $1901.3365$ & $1917.7947$  \\
        \hline
        With $\mu_0$ & $1830.2448$ & $1.0288$ & $1838.2448$ & $1860.1890$  \\ 
        \hline
       
    \end{tabular}
    \caption{Comparison of statistical criteria from two different modeling methods (with and without $\mu_0$).}
    \label{tab:VII}
\end{table}

Table \ref{tab:VII} shows that treating $\mu_0$ as a free parameter in our model results in a lower total chi-squared value ($\chi^2= 1830.2448$) compared to the model without $\mu_0$ ($\chi^2=1895.3365$), indicating a better fit to the data. The reduced chi-squared value for the model with the nuisance parameter ($\chi^2_\nu=1.0288$) is considerably lower than the other one, indicating that this modeling strategy fits the data well within the expected statistical uncertainties.   

 When considering the model selection criteria that include complexity penalties, the model with $\mu_0$ continues to overperform the model without $\mu_0$. The AIC (1838.2448) and BIC (1860.1890) values are significantly lower in the case where we use $\mu_0$ as a free parameter. This reduction reflects once again an improved trade-off between goodness of fit and model complexity, indicating that the model with $\mu_0$ offers a more statistically robust description of combined observational data sets. Also, the differences of AIC and BIC between these two methods are $\Delta\mathrm{AIC} >10$ and $\Delta\mathrm{BIC} >10$ \cite{Kenneth, Kass}, which strongly support the inclusion of the nuisance parameter, demonstrating the significant enhancement of the model's ability to explain the data well.

However, the reduction in $\chi^2$ or AIC/BIC values owing to utilizing the nuisance parameter $\mu_0$ as a free parameter does not indicate that the model explains new cosmology, but rather adjusts for a normalization mismatch.

Given that our analysis is conducted within the context of standard general relativity-based $\Lambda$CDM cosmology, it is justifiable to assert that the gravitational wave luminosity distance $d_L^{\mathrm{GW}}(z)$ is identical to the electromagnetic (EM) luminosity distance $d_L^{\mathrm{EM}}(z)$, denoted as $d_L(z)$ (\ref{36}). This equivalence is maintained in general relativity because both gravitational and electromagnetic waves travel along null geodesics within a spacetime characterized by the same metric, unaffected by modifications from exotic physics, such as modified gravity or energy dissipation into additional dimensions \cite{Nishizawa, Belgacem}. Consequently, in the absence of any divergence from general relativity, the conventional expression for luminosity distance applies to both gravitational wave and electromagnetic signals, represented as \cite{Asghari}
\ben 
d^{GW}_L=c(1+z) \int_{0}^{Z} \frac{dz^\prime}{H(z)^\prime} \label{49}
\een
This approach ensures alignment with the cosmic distance–redshift relationship in a spatially flat FLRW universe and is extensively used in both theoretical research and observational data analysis concerning conventional sirens \cite{LIGOScientific2021, Maggiore}. Then, similarly as Eq. (\ref{37}), we can express 
\ben
\frac{dd^{GW}_L(z)}{dz}= \frac{d^{GW}_L(z)}{1+z}+\frac{c(1+z)}{H(z)}. 
\label{50}
\een

Now we construct the plot of gravitational wave luminosity distance $d^{GW}_L(z)$ as a function of redshift $z$ in Fig. (\ref{Fig8}), we have represented both of the theoretical prediction curve, which is plotted using the Eq. (\ref{50}), and the data points with their individual uncertainties. These data points are basically the different $d^{GW}_L(z)$ values of different GW event catalogs like GWTC-1, GWTC-2, GWTC-2.1 etc.,  collected from {\it the Gravitational Wave Open Science Center (GWOSC)} \cite{GWOSC}, which provides publicly available GW data from LIGO/Virgo/KAGRA detectors. 

\begin{figure}[H]
\centering
\includegraphics[width=8cm, height=5cm]{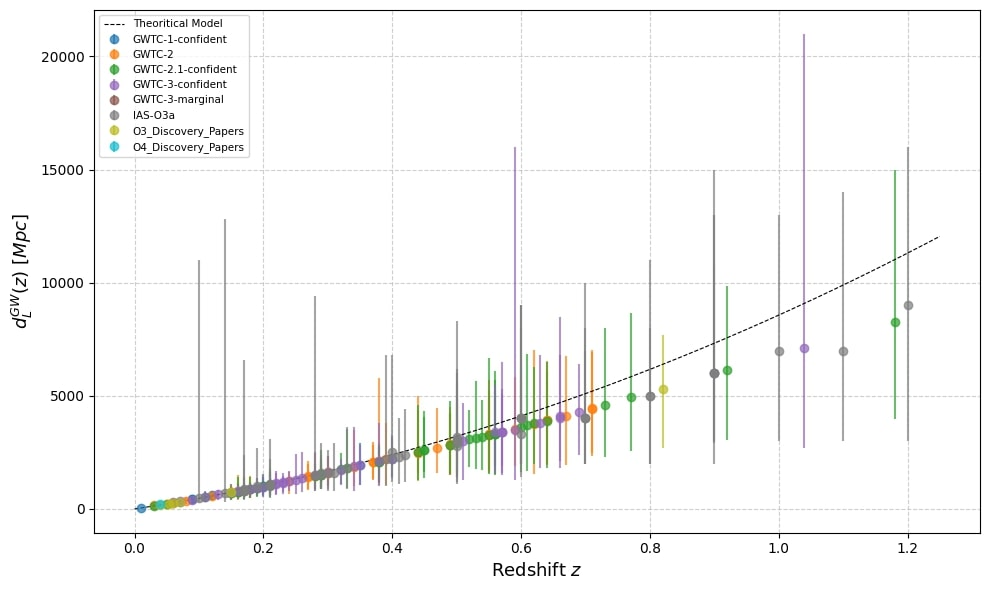}
 \caption{The GW luminosity distance as a function of redshift, from different GW catalogs available in GWOSC \cite{GWOSC}.}\label{Fig10}
\end{figure} 

In Fig. \ref{Fig8}, the dashed curve represents the theoretical prediction for $d^{GW}_L(z)$ under the assumption that GW propagation follows the same geodesics as the electromagnetic radiation in the standard GR-based $\Lambda$CDM cosmology, leading to an expression for the GW luminosity distance that coincides with the EM counterpart, as given in Eq. (\ref{49}). The data points exhibit noticeable scatter and larger uncertainties, particularly at higher redshifts, due to the intrinsic weakness of gravitational wave signals and redshift estimation. Therefore, this plot (Fig. \ref{Fig8}) provides empirical support for the assumption $d^{GW}_L(z)=d^{EM}_L(z)$ \cite{Asghari}, indicating no statistically significant deviation from GR predictions in the propagation of gravitational waves over cosmological distances.  

Eventually, we conclude that we start with the GW-inspired RE (\ref{20}), which may also be expressed within the homogeneous and isotropic FLRW framework in Eq. (\ref{31}), evolved around the standard $\Lambda$CDM model. This modified RE (\ref{20}) emerges naturally from geometric considerations without assuming any particular gravitational theory. By using the perfect fluid representation of the stress-energy tensor (\ref{23}), we construct an equation for the expansion scalar $\Theta$, which, under the FLRW framework, is directly correlated with the Hubble parameter $H(z)$.

The first-order differential equation for the Hubble parameter, $H(z)$ (Eqs. (\ref{31}) or (\ref{32})), governs the expansion history of the universe by converting cosmic time $t$ to redshift space $z$. Solving this equation yields the redshift-dependent Hubble parameter, which can be integrated to compute the luminosity distance $d_L(z)$ or $d^{GW}_L(z)$ for various sources (Eq. (\ref{36}) for EM sources and Eq. (\ref{49}) for GW Standard Sirens).

This chain of derivations from Gw-inspired RE to observable quantities needs no auxiliary assumptions beyond the geometrical structure encoded in the RE and the FLRW metric. All observational quantities, including the Hubble parameter $H(z)$, the luminosity distance $d_L(z)$ and the deceleration parameter $q(z)$, arise directly from the solution of the GW-inspired RE and are constrained using real datasets (Pantheon+, OHD, and BAO). This explicit mapping ensured a transparent theoretical-observational link within a purely geometric framework.

\section{Conclusion}
In this work, we have developed a comprehensive theoretical framework that establishes a novel connection between the Raychaudhuri equation (RE) and gravitational waves within the context of small metric perturbations. Specifically, we derived an essential relation, presented in Eq. (\ref{13}), that connects the Ricci tensor to the perturbed metric tensor ($\Box h_{\alpha \beta}$), utilizing the Lorentz gauge condition to preserve generality and simplify the field equations. Using the relation (\ref{12}), we found the general GW-inspired RE in Eq. (\ref{14}). Furthermore, we have introduced a localized formulation of the energy-momentum distribution in terms of the metric perturbation, encapsulated in Eq. (\ref{17}), which is different from the basic wave Eq. (\ref{18}). This allowed us to reformulate the general GW-inspired RE in Eq. (\ref{14}) into a perturbation-driven version of GW-inspired RE in Eq. (\ref{20}), which reveals the intricate interplay between gravitational wave dynamics and spacetime curvature. The resulting formulation (\ref{14}), or (\ref{20}), not only deepens our understanding of how gravitational waves propagate through spacetime, but also offers new mathematical, physical, and observational insights into the structure and evolution of background spacetime. Starting with the GW-inspired RE and applying it within the homogeneous and isotropic FLRW framework, we established a direct and minimal assumption pathway from the geometric formulation to observational quantities such as $H(z)$, $d_L(z)$, and $q(z)$, ensuring a transparent theoretical-observational connection. 

In this study, we have developed and analyzed our cosmological model under the standard FLRW framework, which reproduces key features of cosmic expansion and exhibits compatibility with current combined observational datasets ({\it Pantheon+Hubble+BAO}). Notably, the inferred value of the Hubble constant $H_0$ from our model, without nuisance parameters, exceeds the estimate from early universe CMB measurement \cite{Planck1, Planck2}, yet remains in close alignment with the local distance ladder result from the SHOES collaboration \cite{Riess}. Furthermore, the sound horizon at the drag epoch $r_d$, inferred from our analysis, is consistent with the latest BAO-based constraints, reinforcing the reliability of the model. 

A key methodological enhancement in our analysis involves the introduction of a nuisance parameter $\mu_0$, designed to marginalize over residual uncertainties related to the absolute magnitude calibration of standard candles (e.g., Type Ia supernovae). This parameter significantly influences the inferred cosmological constraints by decoupling model predictions from systematics in supernova observations. Inclusion of $\mu_{0}$ modifies the theoretical distance modulus, transitioning from Eq. (\ref{35}) to Eq. (\ref{48}), effectively absorbing degeneracies between $H_0$ and $M$ of Type Ia supernovae. As a result, the marginalized posterior distributions reveal a lower value of $H_0$, consistent with the early-universe CMB estimation. In contrast, analyses that exclude yield a higher $H_0$ value consistent with the late-time universe measurements, thus emphasizing the Hubble tension and role of modeling assumptions in shaping cosmological inference. Statistically, the inclusion of $\mu_0$ yields substantially better $\chi^2$, AIC and BIC values, with $\Delta$AIC and $\Delta$BIC both exceeding $10$, strongly favoring the model including $\mu_0$ as a more robust statistical description of the data, though its role remains that of a calibration adjustment rather than a new physical explanation. 

The inclusion of $\mu_0$ not only affects $H_0$ but also yields $r_d=147.67$ Mpc, closely matching the Planck 2018 CMB results $r_d \sim 147-149$ Mpc \cite{Planck1}. The evident interdependence among $H_0$, $r_d$, and $\mu_0$ points to a shared sensitivity to the cosmic expansion history. Since $H_0$ and $r_d$ set the cosmic distance scale, and $\mu_0$ reflects the photometric calibration offsets, their variation highlights the need for rigorous marginalization techniques to isolate cosmological signal from observational artifacts.

From a theoretical perspective, the reformulation of the GW-inspired RE (\ref{20}) in the form of a harmonic oscillator (\ref{47}) within the FLRW background provides a novel lens through which cosmic dynamics can be interpreted. This dynamical analogy offers insights into the behavior of the expansion scalar and its connection to the underlying evolution of the universe. This reinterpretation can be seen alongside existing Hubble tension resolution proposals like Early Dark Energy \cite{Paulin, Karwal} and Interacting Dark Sector models \cite{Ganguly1, Valentino, Pan, He}. Still, our approach achieves a similar type of phenomenological effects entirely within standard GR, without introducing new couplings or fields, via the small metric perturbation.

Here, the GW-inspired RE (\ref{14}), generalized in (\ref{20}) and used inside the FLRW framework as described in Eqs.~(\ref{31})--(\ref{33}), functions as a fundamental tool to analyze cosmological models. Despite our framework diverging from the conventional Friedmann-based methodology, the resultant theoretical curve aligns effectively with OHD, as seen in Fig.~\ref{Fig1}. {\it This agreement supports our model, its assumptions, and the derived cosmological parameters from various data sets, showing that the GW-based Raychaudhuri formulation with small metric perturbations is a viable alternative to conventional cosmological modeling and beyond.} 

In summary, this study demonstrates the efficacy and robustness of the GW-inspired RE framework in cosmological modeling. The use of Bayesian inference techniques, especially via the marginalization of the nuisance parameter $\mu_0$, substantially enhances both the interpretability and statistical robustness of the consequent cosmological constraints. Our findings highlight the intricate relationship between early and late observational data and stress the necessity of using systematics-aware techniques to address key tensions in contemporary cosmology.

Moreover, the GW-inspired RE, as delineated in Eq.~(\ref{14}), facilitates exploration within the realm of higher-dimensional gravity theories, including $f(R)$ gravity~\cite{Sotiriou, Felice, Starobinsky}, $f(R,T)$ gravity~\cite{Harko, Panda3}, etc., providing a foundational framework for future theoretical as well as observational investigations. However, the detailed explanation of those applications is beyond the scope of our present work.\\

{\bf Acknowledgment:}
The authors would like to thank the referees for their illuminating suggestions for improving the paper.  
A.D. would like to thank the Department of Physics, Sister Nivedita University, where the basic portion of the work was done.\\

{\bf Conflicts of interest:} The authors declare no conflicts of interest.\\

{\bf Funding information:} Not available.\\

{\bf Data availability:} The data used in this study are readily accessible from public sources for validation of our model; however, we did not generate any new datasets for this research.\\

{\bf Declaration of competing interest:}
The authors declare that they have no known competing financial interests or personal relationships that could have appeared to influence the work reported in this paper.\\

{\bf Declaration of generative AI in scientific writing:} The authors state that they do not support the use of AI tools to analyze and extract insights from data as part of the study process.\\

\end{document}